\pdfoutput=1
\documentclass[acmtog]{acmart}
\acmSubmissionID{743}
\AtBeginDocument{%
  \providecommand\BibTeX{{%
    \normalfont B\kern-0.5em{\scshape i\kern-0.25em b}\kern-0.8em\TeX}}}

\definecolor{MyRed}{rgb}{0.65,0.07,0.09}

\definecolor{MyGreen}{rgb}{0.18,0.55,0.09}

\usepackage{wrapfig}
\usepackage{diagbox}
\usepackage{verbatim}
\usepackage{amssymb}
\usepackage{latexsym}
\usepackage{lineno}
\usepackage{xspace}
\usepackage{color}
\graphicspath{{figures/}}
\usepackage{amsmath,bm}
\usepackage{psfrag}
\usepackage{mathtools}

\theoremstyle{definition}

\usepackage{multirow, tabularx}
\newcolumntype{Y}{>{\centering\arraybackslash}X}
\usepackage{capt-of}%
\usepackage{array, boldline, makecell, booktabs}

\newcommand{\Mod}[1]{\ (\mathrm{mod}\ #1)}

\usepackage{array}

\usepackage{pifont}

\usepackage{psfrag}
\usepackage{makecell}
\usepackage{algorithm}
\usepackage{threeparttable,booktabs}
\usepackage{graphicx}
\usepackage{subcaption}
\usepackage{flushend}

\usepackage{algpseudocode}

\usepackage{xcolor}
\usepackage{graphicx}

\newcommand{\revv}[1]{\textcolor{black}{#1}}

\newcommand{\WrapFig}[5]
{
\begin{wrapfigure}{r}{{#4}\textwidth}
	\vspace{-.5\baselineskip}
	\centering
	\includegraphics[height={#5}in]{#1}
	\caption{#2}
	\vspace{-.25\baselineskip}
	\label{#3}
\end{wrapfigure}
}

\newif\ifverbose
\verbosetrue

\ifverbose

\newcommand{\vb}[1]{\textcolor{red}{#1}}
\else

\newcommand{\vb}[1]{}
\fi

\usepackage{url}
\usepackage{xcolor}
\definecolor{newcolor}{rgb}{.8,.349,.1}

\usepackage{enumitem}
\usepackage{svg}
\svgpath{{./img/}} 

\setcopyright{rightsretained}
\acmJournal{TOG}
\acmYear{2024} \acmVolume{43} \acmNumber{6} \acmArticle{} \acmMonth{12}\acmDOI{10.1145/3687959}

\begin{document}

\author{Duowen Chen}
\email{dchen322@gatech.edu}
\affiliation{
\institution{Georgia Institute of Technology}
\country{USA}
}

\author{Zhiqi Li}
\email{zli3167@gatech.edu}
\affiliation{
\institution{Georgia Institute of Technology}
\country{USA}
}

\author{Junwei Zhou}
\email{zjw330501@gmail.com}
\affiliation{
\institution{Purdue University},
\institution{University of Michigan}
\country{USA}
}

\author{Fan Feng}
\email{Fan.Feng.GR@dartmouth.edu}
\affiliation{
\institution{Dartmouth College}
\country{USA}
}

\author{Tao Du}
\email{taodu.eecs@gmail.com}
\affiliation{
\institution{Tsinghua University},
\institution{Shanghai Qi Zhi Institute}
\country{China}
}

\author{Bo Zhu}
\email{bo.zhu@gatech.edu}
\affiliation{
\institution{Georgia Institute of Technology}
\country{USA}
}

\title{Solid-Fluid Interaction on Particle Flow Maps}

\begin{abstract}
We propose a novel solid-fluid interaction method for coupling elastic solids with impulse flow maps. Our key idea is to unify the representation of fluid and solid components as particle flow maps with different lengths and dynamics. The solid-fluid coupling is enabled by implementing two novel mechanisms: first, we developed an impulse-to-velocity transfer mechanism to unify the exchanged physical quantities; second, we devised a particle path integral mechanism to accumulate coupling forces along each flow-map trajectory. Our framework integrates these two mechanisms into an Eulerian-Lagrangian impulse fluid simulator to accommodate traditional coupling models, exemplified by the Material Point Method (MPM) and Immersed Boundary Method (IBM), within a particle flow map framework. We demonstrate our method's efficacy by simulating solid-fluid interactions exhibiting strong vortical dynamics, including various vortex shedding and interaction examples across swimming, falling, breezing, and combustion.
\end{abstract}


%
%


\begin{teaserfigure}
 \centering
 \includegraphics[width=.99\textwidth]{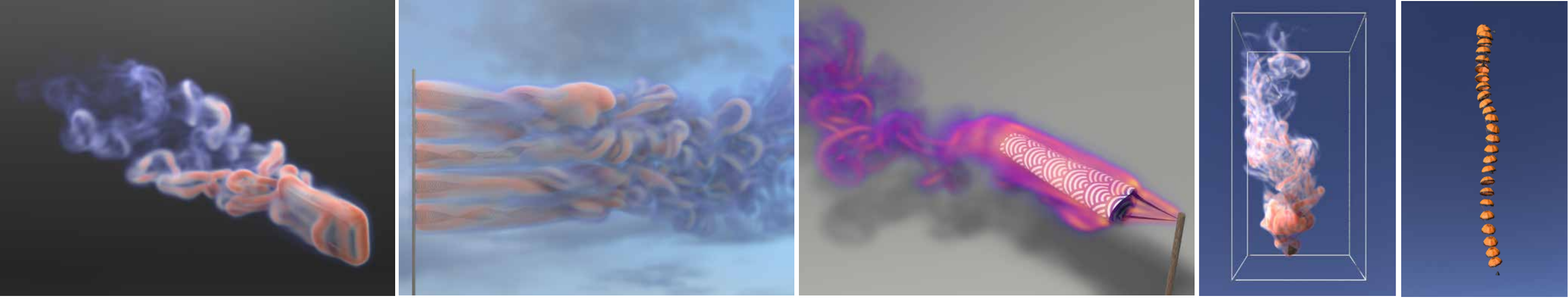}
 \caption{We demonstrate our method's efficacy with various examples of fluid-solid interaction, including a swimming fish, long silk flags, a Koinobori, and a falling parachute along with its trajectory (from left to right), all exhibiting strong vortex dynamics and solid-vortex interactions.}
 \label{fig:teaser}
\end{teaserfigure}

\maketitle

\section{Introduction}
In recent years, flow map methods have attracted increasing attention in computer graphics \cite{qu2019efficient, nabizadeh2022covector} and computational physics \cite{rycroft2020reference}. Unlike traditional approaches that advect physical quantities with flow velocity at each timestep, flow maps establish a (typically bidirectional) point-to-point correspondence between the initial frame and the current frame to transport quantities, achieving impressive long-range advection accuracy through the use of additional data structures such as neural buffers \cite{deng2023fluid} or particles \cite{li2024lagrangian, zhou2024eulerianlagrangian}. One of the most significant advantages of using a flow map is its ability to preserve vortical structures during their creation and evolution, which makes flow maps particularly well-suited for modeling solid-fluid interactions, especially those dominated by vorticity produced around moving boundaries of solids immersed in turbulent fluid environments. Such vortex-solid phenomena are ubiquitous and can be observed in examples like fish swimming, birds gliding, parachutes descending, and wind blowing through thin cloth and hair. 

However, despite the burgeoning literature on using flow map methods to solve fluid flow problems, the study of solid-fluid interactions on flow maps remains sparse. In computer graphics, there has been no previous study on this problem. In computational physics, several prior works (e.g., see \cite{wang2022incompressible,rycroft2020reference}) have developed frameworks to solve solid-fluid interactions based on reference maps within an Eulerian setting, focusing on devising an Eulerian solid framework to fit into the fluid flow map. The primary coupling mechanism in these approaches predominantly relies on creating a narrow band around the solid boundary and mixing stresses between solid and fluid using a Heaviside blending function. This method is highly parameter-sensitive and could produce non-physical results if the narrow band size and the blending function are improperly chosen.

Why do the various classical solid-fluid coupling strategies (e.g., MPM \cite{jiang2016material}, IBM \cite{peskin2002immersed}, variational \cite{batty2007fast}, monolithic \cite{robinson2008two}, etc.) not fit within the flow-map framework? We speculate three potential reasons:
(1) \textit{Flow-map coupling requires a unified representation of both solid and fluid}. In other words, the solid representation and discretization must be identical to their fluid counterparts, such as a flow map with the same initial and final time stamps defined on the same Eulerian grid, which significantly limits the scope of solid models that can be chosen to accommodate complex solid-fluid interactions. In particular, the significance of devising a long-range flow map model for solid simulation remains unclear due to the less connected nature between flow advection and solid dynamics.   
(2) \textit{Flow-map coupling requires exchangeable physical quantities between solid and fluid}. Though this was not a problem in conventional solid-fluid interaction frameworks (e.g., exchanging velocity or momentum via G2P or P2G operations in traditional MPM), modern flow map methods typically evolve gauge variables (e.g., impulse \cite{cortez1996impulse}, vorticity \cite{cottet2000vortex}, and other gauges \cite{saye2016interfacial, saye2017implicit}) instead of fluid velocities, which cannot be directly operated with solid velocities. For instance, we cannot naively conduct a P2G operation across the solid-fluid interface with fluid particles carrying impulses and solid particles carrying velocities. 
(3) \textit{Adding an external force to a flow-map model remains an open problem}. Although simple forces such as gravity can be incorporated into existing flow-map models, their physical accuracy is less grounded. Local forces, such as momentum exchange, remain unclear regarding how they should be transferred from solid to fluid in a flow-map system.

We propose a novel solid-fluid interaction framework based on flow map models by addressing the abovementioned challenges. Our key idea is to model both solid and fluid as a unified forward flow map on particles: each fluid particle represents a long flow map governed by impulse fluid dynamics, while each solid particle represents a short flow map governed by elastic solid dynamics. Specifically, we restrict the solid flow map to a single time step to adapt an arbitrary conventional solid simulation model (e.g., MPM or XPBD). The fluid and solid flow maps are coupled based on two key mechanisms: (1) we implement an impulse-to-velocity transfer mechanism to unify the physical quantities exchanged between solid and fluid particles; (2) we implement a particle path integral mechanism to accurately accumulate both pressure and coupling forces along each flow-map trajectory. The combination of these two mechanisms, in conjunction with the standard particle-grid operations and incompressibility projections, synergistically enables a versatile coupling framework to exchange information between particle flow maps with different lengths and governing physics, which further accommodates the adaptation of various traditional coupling models into flow map methods. In our implementation, we demonstrate two examples of MPM and IBM coupling by integrating both into a hybrid Eulerian-Lagrangian fluid simulator on particle flow maps. Thanks to the inherent advantage of preserving vortical structures in our flow map model, these flow-map-enhanced coupling systems produce vortex-solid interaction simulations that outperform traditional methods in terms of both physical accuracy and visual complexity. In our experiments, we implemented a diverse set of benchmark tests and simulation examples, ranging from leaves falling and fish swimming to complex vortex shedding behind cloth, hair, and combustion processes, demonstrating our framework's versatility and efficacy in tackling complex vortex-object interaction simulations.

We summarize our main contributions as follows:
\begin{itemize}
    \item A unified particle flow-map representation with different lengths and governing equations for fluid and solid;
    \item A reformulated impulse gauge fluid model to enable solid-fluid momentum exchange on particles;
    \item A path integral approach on particle flow maps to accumulate coupling forces;
    \item A versatile framework to accommodate traditional solid-fluid coupling mechanisms on flow-map models.  
\end{itemize}
\section{Related Work}

\subsection{Flow Map \& Impulse Fluid}
Initially known as the method of characteristic mapping (MCM), the concept of flow maps was first introduced by \citet{wiggert1976numerical}. By reducing the diffusion error caused by semi-Lagrangian advection, various attempts in the graphics community were made to adapt this method in fluid simulation \cite{hachisuka2005combined, sato2017long, sato2018spatially, tessendorf2015advection, qu2019efficient}. In Covector Fluid (CF) \cite{nabizadeh2022covector}, flow maps were first introduced to aid the advection of the covector variable, achieving state-of-the-art vorticity preservation effects. The impulse variable, a form of covector, was first introduced by \citet{buttke1992lagrangian}. By rewriting the incompressible Navier-Stokes Equations through the use of a gauge variable and gauge transformation \cite{oseledets1989new, roberts1972hamiltonian, buttke1992lagrangian, buttke1993velicity}, it allows for the gauge freedom to be designed for specific applications \cite{buttke1993velicity, buttke1993turbulence, weinan2003gauge, cortez1996impulse, summers2000representation, saye2016interfacial, saye2017implicit}. This concept was revisited in computer graphics by \citet{feng2022impulse} and \citet{yang2021clebsch}. Neural Flow Map (NFM) \cite{deng2023fluid} further enhances the flow-map accuracy with a neural buffer. Recently, Particle Flow Map (PFM) \cite{zhou2024eulerianlagrangian} and Impulse PIC (IPIC) \cite{sancho2024impulse} used a hybrid method with particles, and \cite{li2024lagrangian} extended the covector to pure Lagrangian representations.

In \cite{cortez1996impulse, summers2000representation, saye2016interfacial, saye2017implicit}, attempts to couple impulse with solids were made. However, these methods were limited by the need to redesign the gauge variable for different solids \cite{saye2016interfacial, saye2017implicit} and could not be adapted to the advection scheme using flow maps \cite{cortez1996impulse, summers2000representation}. Our method aims to use PFM to design a general solid-fluid coupling scheme for impulse fluids within the flow map framework.

\subsection{Full Eulerian Coupling}
For full Eulerian methods, computation time benefits arise from both solid and fluid being treated on a single fixed background grid. Such methods include the deformation gradient-based method \cite{liu2001eulerian} and initial point set (IPS) \cite{dunne2006eulerian}. Recently, the reference map technique (RMT) \cite{kamrin2009eulerian, kamrin2012reference, rycroft2020reference} has attracted wide attention and was later extended to couple rigid bodies with fluid \cite{wang2022incompressible}. As for pure Eulerian treatment in graphics, \citet{teng2016eulerian} allows for larger time steps in pure Eulerian solid-fluid coupling by setting up a semi-implicit coupling system. Concerning flow maps, by noting the correspondence between the flow map and the deformation gradient used in the elastic solid simulation, RMT tries to incorporate them together on Eulerian mesh and bridge the two with a Heaviside function but is limited by the requirement of a narrowband blending scheme.

\subsection{Full Lagrangian Coupling} 
Full Lagrangian methods use Lagrangian elements in both the fluid and solid domains. Representatives of this method in computational physics include particle FEM methods \cite{idelsohn2008unified, cremonesi2020state, becker2015unified} and, later, the combination of MPM and FEM \cite{lian2011femp, lian2014coupling, lian2011coupling, lian2012adaptive}. As for research in graphics, it was initially explored in \cite{muller2004point, keiser2005unified}. Later, \citet{klingner2006fluid} proposed using a body-confronting mesh for coupling. Subsequent works proposed a unified framework representing both solid and fluid with Lagrangian elements \cite{clausen2013simulating}. Coupling SPH with deformable has also been explored \cite{solenthaler2007unified, akinci2013coupling} but is limited to relatively simple simulation settings. In \cite{akbay2018extended}, authors proposed an extended partition method (XPM) and demonstrate this using Lagrangian solid coupling with an Eulerian fluid solver on the grid and a Lagrangian fluid solver like SPH.

\subsection{Mixed Lagrangian-Eulerian Coupling}
In mixed Lagrangian-Eulerian mesh methods, solids are represented by Lagrangian markers coupled with fixed Eulerian background meshes. Representatives of such methods include the immersed boundary method (IBM) \cite{peskin1972flow, peskin2002immersed, mori2008implicit, huang2009immersed}. Based on IBM, the immersed finite element method \cite{liu2006immersed, liu2007mathematical, shimada2022eulerian}, immersed interface method \cite{zhao2008fixed}, and immersed continuum method \cite{wang2007iterative, wang2006immersed} were proposed. Other improvements on IBM methods have been made, such as monolithic projection methods \cite{wang2020monolithic} and combining marker particles with finite volume expression \cite{shimada2022eulerian}. See \cite{huang2019recent} for more details.
In the graphics community, since the pioneering work of \citet{carlson2004rigid}, \citet{genevaux2003simulating}, and \citet{guendelman2005coupling}, various works in this direction have been explored. \citet{batty2007fast} and \citet{ng2009efficient} treated coupling as an energy minimization form, \citet{robinson2008two}, \citet{robinson2009accurate} and \citet{robinson2011symmetric} perform implicit coupling and eables the free-slip boundary conditions. \citet{zarifi2017positive}, \citet{takahashi2020monolith} and \citet{takahashi2022elastomonolith} use the cut-cell technique combined with a properly constrained global system for further improvement. As for hybrid methods like MPM \cite{stomakhin2013material}, it comes naturally with non-slip boundary conditions for coupling, and further research extends it to enable rigid body coupling \cite{hu2018moving} and the free-slip boundary conditions \cite{fang2020iq}.
Exploration of hybrid methods in incorporating flow maps was present in \cite{shimada2021eulerian}. Authors improved RMT to a hybrid method with marker particles to better track the solid interface. However, the main problem of such a method is that fluid does not directly benefit from the advection using a flow map and, therefore, cannot achieve simulation quality as shown in impulse-based fluid methods. 

\subsection{Coupling with Thin Structures}
Using the coupling techniques mentioned above, different scales of coupling phenomena have also been studied. In particular, coupling with thin structures is of interest and examples include coupling hair-fluid coupling \cite{fei2017multi}, fabric-fluid coupling \cite{fei2018multi}, coupling fabric with non-Newtonian fluid \cite{fei2019multi}, coupling parachute/cloth with fluid \cite{wang2020monolithic}, insect flying and fish swimming \cite{tian2014fluid, cui2018sharp, borazjani2010role}, coupling uniform flow with flags \cite{uddin2013interaction, wang2019numerical}, coupling free-surface water with thin shells \cite{robinson2008two} and also combustion between fire and paper or cloth \cite{losasso2006melting}

\newcolumntype{z}{X}
\newcolumntype{s}{>{\hsize=.25\hsize}X}
\begin{table}[t]
\centering\small
\begin{tabularx}{0.47\textwidth}{scz}
\hlineB{3}
Notation & Type & Definition\\
\hlineB{2.5}
\hspace{12pt}$\bm *^f$ & vector/matrix & fluid property\\
\hlineB{1}
\hspace{12pt}$\bm *^s$ & vector/matrix & solid property\\
\hlineB{1}
\hspace{12pt}$\bm *^n$ & vector/matrix & fluid properties near solid (only used in MPM coupling)\\
\hlineB{1}
\hspace{12pt}$\bm *_t$ & scalar/vector/matrix & quantities evaluated at time $t$\\
\hlineB{1}
\hspace{12pt}$\bm x$ & vector & particle / mesh vertices location\\
\hlineB{1}
\hspace{4pt} $\mathcal{F}_{[c, a]}$ & matrix & Forward Jacobian from $c$ to $a$\\
\hlineB{1}
\hspace{4pt} $\mathcal{T}_{[a, c]}$ & matrix & Backward Jacobian from $a$ to $c$\\
\hlineB{1}
\hspace{12pt}$\mathbb{C}$ & function & constraint in XPBD simulation\\
\hlineB{1}
\hspace{12pt}$\bm{u}$ & vector & velocity\\
\hlineB{1}
\hspace{12pt}$\nabla\bm{u}$ & matrix & velocity gradient\\
\hlineB{1}
\hspace{12pt}$\bm{m}$ & vector & impulse\\
\hlineB{1}
\hspace{12pt}$\bm f$ & vector & force \\
\hlineB{1}
\hspace{12pt}$p$ & scalar & pressure \\
\hlineB{1}
\hspace{12pt}$\bm \Lambda$ & vector & pressure correction buffer \\
\hlineB{1}
\hspace{12pt}$\bm \Upsilon$ & vector & external force buffer \\
\hlineB{1}
\hspace{12pt}$n$ & scalar & reinitialization steps\\
\hlineB{3}
\end{tabularx}
\vspace{5pt}
\caption{Summary of important notations used in the paper.}
\label{tab: notation_table}
\end{table}

\section{Physical Model}

\paragraph{Naming Convention}
We will adhere to the naming conventions in Table~\ref{tab: notation_table}. Specifically, we will use superscripts for the type a quantity belongs to. For example, $*^f$ denotes fluid-related quantities, and $*^s$ denotes solids-related ones. We will use subscripts to indicate the evaluation time of a quantity, such as $*_t$ for values evaluated at time $t$. Similarly, $*_{[a, c]}$ represents a time interval from $a$ to $c$ and is used in flow map notations to indicate the duration over which the mapping occurs, with $a$ and $c$ as starting and ending time. 

\subsection{Governing Equations}
We lay out the physical model of the solid-fluid coupling system on impulse variable:
\begin{equation}
\begin{dcases}
 \frac{D \bm m }{D t} =-\left(\bm \nabla \bm u\right)^{T}\,\bm m,\\
\frac{1}{\rho^f}\nabla^2 \varphi = -\bm \nabla \cdot \bm m,\\
\bm u = \bm m - \frac{1}{\rho^f}\bm \nabla \varphi,\\
\rho^s\frac{D\bm v}{Dt} = \nabla \cdot \sigma,\\
\bm u = \bm v, \quad \mathbf{x}\in \partial \Omega^{s, f}.
\end{dcases}
\label{eq:m_NS}
\end{equation}
where $\bm m$ and $\bm u$ being fluid impulse and fluid velocity, and $\varphi$ an intermediate variable used only for projecting $\bm m$ to the divergence-free $\bm u$. Here, $\bm v$ represents solid velocity, and $\sigma$ is the elastic stress tensor. Here we use $\partial \Omega^{s, f}$ to represent the interface between fluid and solid. The first three equations describe the fluid momentum (first row) and incompressibility (second and third rows), the fourth equation describes the solid, and the fifth equation features the non-slip boundary conditions on the solid-fluid interface. 

\subsection{Flow Map}
Flow map defines a bidirectional mapping $\phi$ and $\psi$ between material space $\bm X$ and world space $\bm x$. The quantity defining infinitesimal change in each space resulting in changes in another can also be described using their Jacobians $\mathcal{F}$ and $\mathcal{T}$. 

Specifically, we define the forward flow map \(\phi(\cdot, t)\) as a function of space and time, mapping the initial position of a particle at time \(0\) to its position at a subsequent time \(t\), and backward flow map \(\psi(\cdot, t)\), as the mapping from time \(t\) back to time \(0\). We can define the forward flow map as:
\begin{equation}
    \begin{dcases}
        \frac{\partial \phi\left(\bm{X}, \tau\right)}{\partial \tau} = \bm{u}\left(\phi\left(\bm{X}, \tau\right), \tau\right), \\
        \phi\left(\bm{X}, 0\right) = \bm{X}, \\
        \phi\left(\bm{X}, t\right) = \bm{x},
    \end{dcases}\\
    \label{eq:flow_map}
\end{equation}
Similarly, we can define the backward flow map as:
\begin{equation}
    \begin{dcases}
        \frac{\partial \psi\left(\bm{x}, \tau\right)}{\partial \tau} = \bm{u}\left(\psi\left(\bm{x}, \tau\right), \tau\right), \\
        \psi\left(\bm{x}, t\right) = \bm{x}, \\
        \psi\left(\bm{x}, 0\right) = \bm{X},
    \end{dcases}
\end{equation}
Subsequently, \(\mathcal{F}\) and \(\mathcal{T}\), the Jacobians of \(\phi\) and \(\psi\), can be calulated by \(\mathcal{F} := \frac{\partial \phi}{\partial \bm{X}}\), \(\mathcal{T} := \frac{\partial \psi}{\partial \bm{x}}\) and satisfies the following advection equations with a velocity field $\bm u$:
\begin{equation}
    \begin{dcases}
    \frac{D \mathcal{F}}{D t}=\bm \nabla\bm{u}\mathcal{F},\\
    \frac{D \mathcal{T}}{D t}=-\mathcal{T}\bm \nabla\bm{u}. \label{eq:T_mapping}
    \end{dcases}
\end{equation}
Here, \(\frac{D(\cdot)}{Dt}\) represents the material derivative, which describes the rate of change of the Jacobians moving with the particle along its flow map's trajectory. 
As shown in previous works \cite{cortez1996impulse,nabizadeh2022covector,deng2023fluid}, fluid impulse can be transported via a bidirectional flow map with the defined Jacobians as:
\begin{equation}
\begin{dcases}
    \bm m(x,t)&=\mathcal{T}^{T}\,\bm{m}(\psi(\bm x),0),\\
    \bm{m}(X, 0) &= \mathcal{F}^T\,\bm m(\phi(\bm X),t).
\end{dcases}
\end{equation}

\paragraph{Particle Flow Map}
\WrapFig{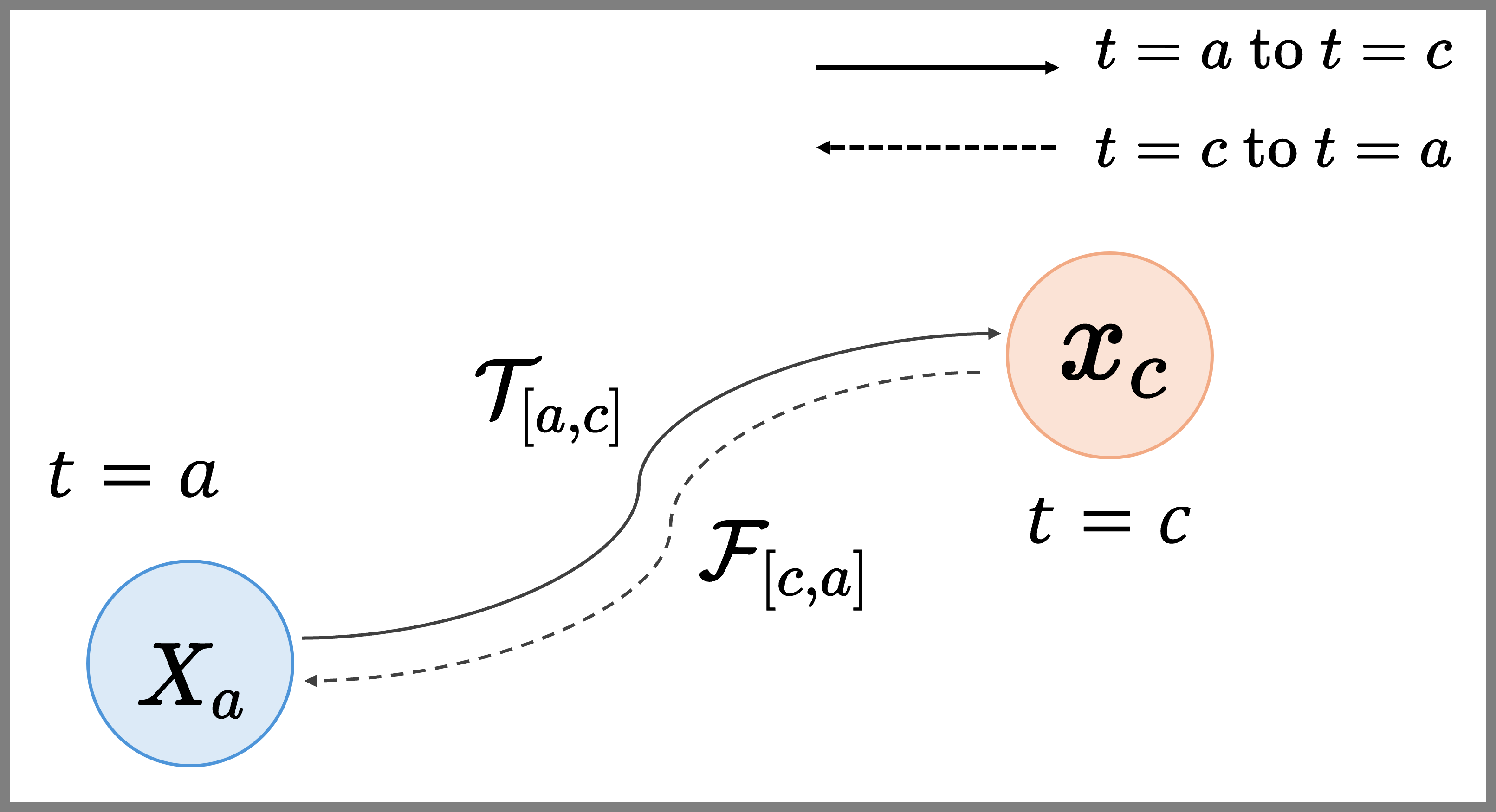}{Particle Flow Map starting from time $a$ and ending at time $c$. }{fig:pfm}{.265}{0.95}
By observing that a particle trajectory can naturally characterize a bidirectional flow map, a particle flow map method was proposed in \cite{zhou2024eulerianlagrangian} by combining affine Particle-in-Cell \cite{jiang2015affine}, covector/impulse fluid \cite{nabizadeh2022covector}, and bidirectional flow map \cite{deng2023fluid} as depicted in Figure~\ref{fig:pfm}. Under a particle perspective, fluid quantities such as $\bm m$, $\mathcal{T}$, and $\mathcal{F}$ are carried on Lagrangian particles. Its trajectory, originating from time $a$ and culminating at time $c$, characterizes both the forward map ${\phi}_{[c, a]}$ and the backward map ${\psi}_{[a, c]}$. 
The Jacobians $\mathcal{F}$ and $\mathcal{T}$ are evolved along the trajectory and facilitate the mapping between the initial and current time frames.


\section{Coupling System}
\subsection{Solid-Fluid Coupling on Flow Maps}
With the flow map theory in hand, we next consider a simple solid-fluid interaction problem. 
Suppose we have an immersed solid $\Omega^s$ in fluid $\Omega^f$ and their interface is denoted as $\partial \Omega^{s, f}$. We denote the force acting on fluid from solid as $\bm f^{s\rightarrow f}$ and the force on solid from the fluid as $\bm f^{f\rightarrow s} = -\bm f^{s\rightarrow f}$. From Euler's equation, the advection equation of impulse variable and the definition of fluid impulse, we can describe the evolution of fluid velocity and impulse in a coupling system as:
\begin{equation}
    \begin{dcases}
        \frac{D \bm u}{Dt} = -\nabla p + \bm f^{s\rightarrow f},\\
        \frac{D\bm m}{Dt}=-(\nabla \bm u)^T \bm m\\
        \bm u = \bm m - \bm q.\\
    \end{dcases}
    \label{eq:sfi_u}
\end{equation}
In the original impulse method, since there were no external forces, \revv{\( \bm q = \nabla \varphi \) for \( \bm q \)} only needed to include pressure. In fluid-structure interaction problems, \( \bm q \) needs to include external forces such as those exerted by the solid, hence the expression for \revv{\( \bm q \)} needs to be rederived. For simplicity in notation, we use \( \bm{f} := f^{s \rightarrow f} \). From Eq. ~\ref{eq:sfi_u}, we can obtain:
\revv{
\begin{equation}
    \begin{split}
        &\frac{D\bm u}{Dt} = \frac{D \bm m}{D t} - \frac{D \bm q}{D t},\\
        &-\nabla p + \bm f + \frac{D \bm q}{D t} + (\nabla \bm u)^T\bm m = 0,\\
        &-\nabla p + \bm f + \frac{D \bm q}{D t} + (\nabla \bm u)^T\bm u + (\nabla \bm u)^T\bm q = 0,\\
        &\frac{D \bm q}{D t} + (\nabla \bm u)^T\bm q + \nabla(\frac{1}{2}|\bm u|^2) - \nabla p + \bm f = 0,
    \end{split} 
\end{equation}
Here, we used the vector identity $(\nabla \bm u)^T\bm u = \frac{1}{2} |\bm u|^2$. By applying Eq.~\ref{eq:T_mapping} and flow map identity $\mathcal{F}\mathcal{T} = \mathcal{I}$, we have:
\begin{equation}
    \begin{split}
        &\frac{D \bm q}{D t} + \mathcal{T}^T \frac{D \mathcal{F}^T}{D t}\bm q + [\nabla(\frac{1}{2}|\bm u|^2) - \nabla p + \bm f] = 0,\\
        &\frac{\mathcal{F}^T D\bm q}{D t} + \frac{D \mathcal{F}^T}{D t}\bm q + \mathcal{F}^T[\nabla(\frac{1}{2}|\bm u|^2) - \nabla p + \bm f] = 0,\\
        &\frac{D\mathcal{F}^T\bm q}{D t} = \mathcal{F}^T[\nabla p - \nabla(\frac{1}{2}|\bm u|^2)]  - \mathcal{F}^T\bm f.
    \end{split}
    \label{eq:buffer_continue}
\end{equation}
}

Assume the flow map starts from time $a$ and ends at time $c$, by integrating both sides of the above equation and multiplying with $\mathcal{T}^T$, we have the following equation at time $c$:
\begin{equation}
    \bm q_c = \mathcal{T}_{[a, c]}^T[\int_a^c \mathcal{F}^T_{[\tau, a]} (\nabla p_{\tau} - \nabla\frac{1}{2}\bm |u_{\tau}|^2)d\tau -\int_a^c \mathcal{F}^T_{[\tau, a]}\bm f_{\tau} d\tau].
\end{equation}
Now we have rederived an expression for $\bm q$ at time $c$ with the presence of external forces and by applying the equation $\bm u = \bm m - \bm q$, we obtain the expression of fluid velocity at the end of the flow map as:

\begin{equation}
    \begin{split}
        \bm u_c &= \underbrace{\mathcal{T}_{[a, c]}^T\bm m_a}_{\text{Impulse Mapping}}- \underbrace{\mathcal{T}_{[a, c]}^T\int_a^c \mathcal{F}^T_{[\tau, a]} (\nabla p_{\tau} - \nabla\frac{1}{2}\bm |u_{\tau}|^2)d\tau}_{\text{Projection}} \\
        &+ \boxed{\underbrace{\mathcal{T}_{[a, c]}^T\int_a^c \mathcal{F}^T_{[\tau, a]} \bm f_{\tau} d\tau)}_{\text{Coupling Force Integral}}},
    \end{split}
    \label{eq:sfi_u_continuous}
\end{equation}
where $\tau$ denotes the intermediate time instant along the trajectory of the flow map.


\begin{wrapfigure}[14]{r}{0pt}
\centering
\includegraphics[width=0.25\textwidth]{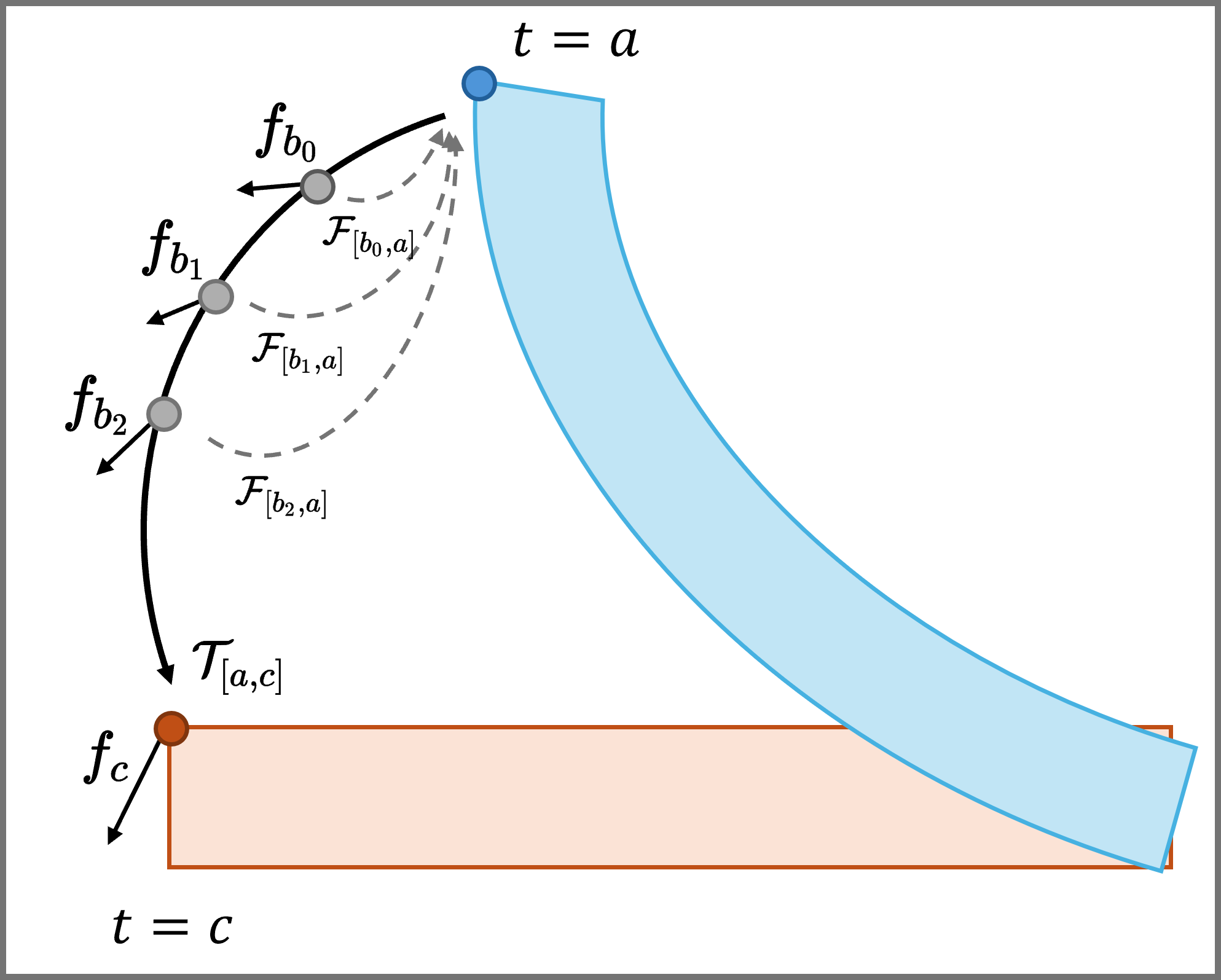}
\caption{Illustration for mapping coupling force from current to initial frame.}\label{fig:coupling_solid}
\end{wrapfigure}
The three terms on the right-hand side of Eq. \ref{eq:sfi_u_continuous} indicate the flow map of fluid impulse, the projection, and the coupling force integral. For an impulse flow-map system without solid-fluid coupling, only the first two terms contribute to calculating velocity in each time step. For a solid-fluid coupling system, the third term is essential for evaluating the coupling effects from solid to fluid. As shown in Figure~\ref{fig:coupling_solid}, the calculation of the coupling force integral can be comprehended as integrating the coupling force with flow maps along the entire trajectory. For each time instant, the integral is calculated by conducting a backward map from the current to the initial time and then conducting a forward map from the initial to the end. 

For the solid part, we will have:
\begin{equation}
    \bm v_c = \bm v_a + \int_a^c \nabla \cdot \sigma_{\tau}(\mathcal{F}_{\tau}) -\bm f_{\tau} d\tau.
\end{equation}
The solid velocity at time $c$ is the time integral of both its elastic force and the fluid force from time $a$ to time $c$. 
The velocity on the solid-fluid boundary at time $c$ must satisfy
$
    \bm u_c = \bm v_c.
$ In order to utilize existing solid simulation methods, in practice we use one-step advection to simulate solids in place of using a long-range mapping.

\subsection{Impulse-To-Velocity Coupling}
\label{sec:imp_to_vel}

\label{sec:impulse_to_vel_numerical}
In this section, we introduce how we solve Eq. \ref{eq:buffer_continue} and \ref{eq:sfi_u_continuous} with a set of reformulated flow-map equations. Our key idea is to convert the impulse-form flow map to the velocity-form flow map to solve the coupled system, allowing a direct coupling implementation based on the force exchange within a single time step. The mathematical idea of our impulse-to-velocity transfer was motivated by \cite{li2024lagrangian}, with our particular focus on deriving the formula in an Eulerian-Lagrangian setting for tackling solid-fluid coupling. 

\paragraph{Particle Buffer}
Our coupling scheme performs in the velocity domain as shown in Eq.~\ref{eq:sfi_u_continuous}. The calculation of two integrals is needed to convert the flow map advected impulse to velocity. We name the two buffers for computing integrals in Eq.~\ref{eq:sfi_u_continuous} as pressure correction buffer $\bm \Lambda$ and coupling force buffer $\bm \Upsilon$:
\begin{equation}
    \begin{dcases}
        \bm \Lambda_c &= \int_a^{c} \mathcal{F}^T_{[\tau, a]} \nabla (p_{\tau} - \frac{1}{2}|\bm{u_{\tau}}|^2)d\tau,\\
        \bm \Upsilon_c &= \int_a^{c} \mathcal{F}^T_{[\tau, a]} \bm f_{\tau} d\tau.
    \end{dcases}
    \label{eq:external_force}
\end{equation}
Grid-based flow map methods cannot easily calculate path integrals starting from a random point within the computational domain, as seen in \cite{sato2017long, deng2023fluid} due to the need for a storage buffer for velocity fields. Meanwhile, as shown in \cite{li2024lagrangian,zhou2024eulerianlagrangian}, particles can easily accumulate over their trajectories. Therefore, we employed a particle flow map framework to integrate the force model into impulse fluid simulation. 
Specifically, we carry individual flow map gradients $\mathcal{F}$ and $\mathcal{T}$ and initial impulse variables $\bm m_a$ on fluid particles. Such design allows the particle buffers $\bm \Lambda$ and $\bm \Upsilon$ to be directly carried and calculated on particles at each time step, making using the particle buffers possible. 
\paragraph{Impulse-to-Velocity Conversion}
To use the particle buffers getting $\bm u_c$ from $\bm m_c$ in simulation, we need to discretize Eq.~\ref{eq:sfi_u_continuous} and Eq.~\ref{eq:external_force} that forms a single step update equation from time $b = c -\Delta t$ to $c$. However, in the definition of $\bm m$, $\bm u$ is a divergence-free velocity field. Therefore, Eq.~\ref{eq:sfi_u_continuous} contains pressure at time $c$ that's unknown and waiting to be solved by Poisson projection. Similarly, $\frac{1}{2}\bm u_c^2$ is unknown and waiting to be solved. Hence, the actual divergent velocity field \revv{$\bm u_c^*$} at time $c$ needs to eliminate the pressure component from Eq.~\ref{eq:sfi_u_continuous} and we use $\frac{1}{2}\bm u_b^2$ to approximate the original expression for $\frac{1}{2}\bm u_c^2$. And, in practice, $\frac{1}{2}\bm u_{\text{mid}}^2$ is used because of our adaptation of mid-point velocity approximation. Hence, Eq.~\ref{eq:sfi_u_continuous} can now be written as:
\begin{equation}
    \begin{aligned}
        \bm u_c^* = \mathcal{T}_{[a, c]}^T(\bm m_a) -\mathcal{T}_{[a, c]}^T(\mathcal{F}_{[b, a]}^T\bm q_b) +\Delta t \bm f_c + \Delta t(\nabla\frac{1}{2}|\bm u_b|^2)
    \end{aligned}
\end{equation}
for a single timestep update from $t=b$ to $t=c$. For deriving update equation for $\bm \Lambda$ and $\bm \Upsilon$, we perform discretization on Eq.~\ref{eq:external_force} and derive:
\begin{equation}
\begin{split}
        \bm \Lambda_c &= \bm \Lambda_b + \mathcal{F}^T_{[c, a]}\Delta t(\nabla p_c - \nabla\frac{1}{2}|\bm u_c|^2)\\
        \bm \Upsilon_c &= \bm \Upsilon_b + \mathcal{F}^T_{[c, a]}\Delta t\bm f_c
\end{split}
\end{equation}

\begin{figure}
    \centering
    \includegraphics[width=0.99\columnwidth]{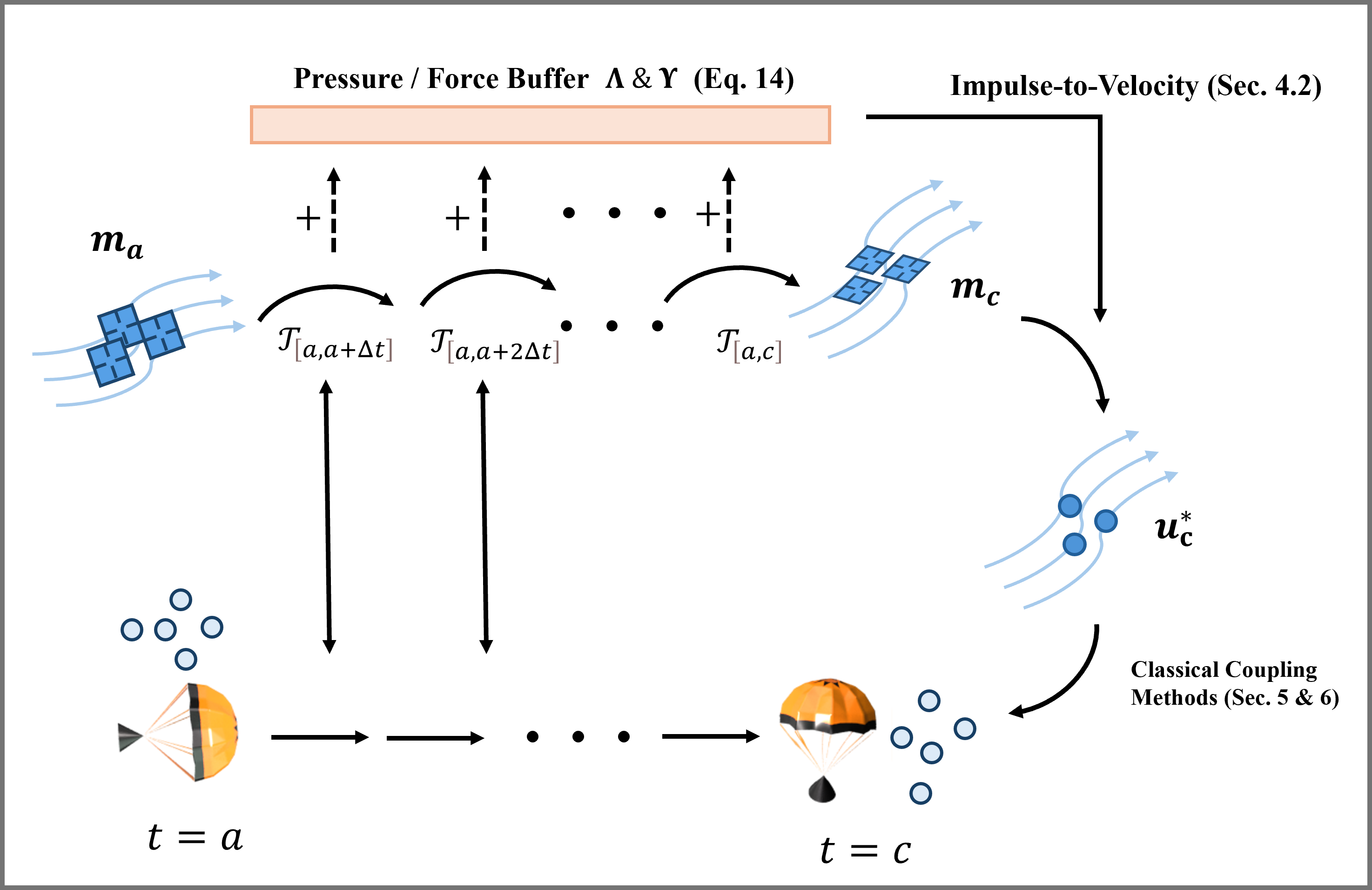}
    \caption{\revv{Illustration of the key idea of our method where we convert impulse to velocity using our particle buffers $\bm \Lambda$ and $\bm \Upsilon$. The problem of direct coupling between impulse and velocity is shown in Section~\ref{sec:ablation}. Our whole pipeline is shown in Section 5 and the adaptation of our method to classical coupling methods is shown in Section~\ref{sec:ibm} and \ref{sec:mpm}.}}
    \label{fig:illu}
\end{figure}

\paragraph{Reformulated Flow-Map Equations}
Now we have the system of equations for updating $\bm m_c$, $\bm u_c^*$, $\bm \Lambda_c$, and $\bm \Upsilon_c$ and we summerize them as follows:
\revv{
\begin{equation}
    \begin{dcases}
        \bm m_c = \mathcal{T}_{[a, c]}^T\bm m_a,\\
        \bm u_c^* = \bm{m}_c - \mathcal{T}_{[a, c]}^T (\bm \Lambda_b - \bm \Upsilon_b) + \nabla \frac{1}{2}|\bm{u}_b|^2 \Delta t + \bm f_c\Delta t,\\
        \bm \Lambda_c = \bm \Lambda_b + \mathcal{F}^T_{[c, a]}\Delta t(\nabla p_c - \nabla\frac{1}{2}|\bm u_c|^2),\\
        \bm \Upsilon_c = \bm \Upsilon_b + \mathcal{F}^T_{[c, a]}\Delta t\bm f_c.
    \end{dcases}
    \label{eq:update_equation}
\end{equation}
}
Notice we use $\bm f_c$ denote the coupling force between fluid and solid at time step $c$, but external forces like gravity $\bm f_c = \rho \bm g$ and viscosity $\bm f_c = \upsilon \Delta \bm{m}_c$ can also be incorporated to $\bm \Upsilon$ in the same way.
Furthermore, in scenarios where non-uniform density is present, $\bm \Lambda_c$ becomes: 
\begin{equation}
    \bm \Lambda_c = \bm \Lambda_b + \mathcal{F}^T_{[c, a]}\Delta t(\frac{1}{\rho}\nabla p_c - \nabla\frac{1}{2}|\bm u_c|^2).
    \label{eq:density}
\end{equation}




\begin{figure*}[t]
 \centering
 \includegraphics[width=.99\textwidth]{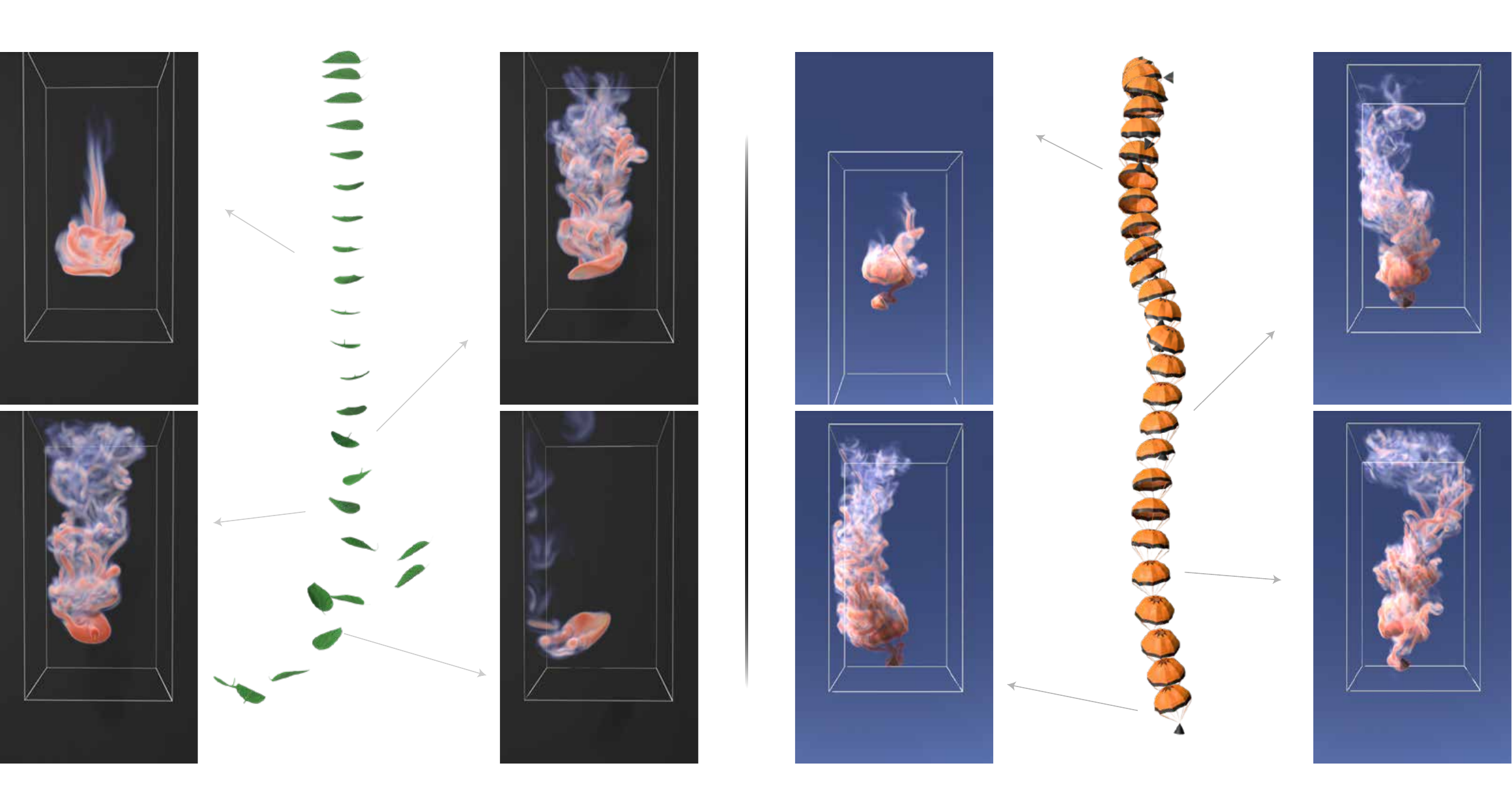}
 \caption{We show the path and vorticity snapshots of free falling leaf and parachute. The falling path of the leaf is shown on the left. We can clearly observe the falling path experienced three stages: acceleration, converging, and chaos which accords with real-life observation. On the right, we show a parachute falling and we observe velocity convergence after the acceleration stage as expected.}
 \label{fig:3d_leaf_para}
\end{figure*}

\paragraph{Calulation of $\bm \Lambda_c$} For evaluating $\bm \Lambda_c$, it directly follows the update equation in Eq.~\ref{eq:update_equation}. After solving the Poisson projection for time $c$, we can obtain ${p_c}$ and $\bm u_c$ on the grid. We then calculate $\nabla (p_{c} - \frac{1}{2}|\bm{u_{c}}|^2)$ on the grid and transfer this quantity to particles using a G2P process and accumulate the value on $\bm \Lambda_c$ by mapping it back to the initial frame $a$ using $\mathcal{F}_{[c, a]}$. The gradient operator can be evaluated as the gradient of the weight kernel used in PFM. Notice that $\rho$ is a material property carried by particles. Therefore, the transition from uniform density to non-uniform density calculation is straightforward.

\paragraph{Calulation of $\bm \Upsilon_c$} As for evaluating $\bm \Upsilon_c$, same procedure is used. At time $c$, we first calculate the coupling force $\bm f_c$ along with optional external forces. Then, using the advected $\mathcal{F}_{[c, a]}$ on particles, we accumulate the forces to $\bm \Upsilon$ by Eq.~\ref{eq:update_equation}.

\paragraph{Calulation of $\bm u_c^*$} Having the two buffers in hand, converting from impulse $\bm m_c$ to divergent velocity $\bm u_c^*$ become straight forward. We use $\mathcal{T}_{[a, c]}^T\bm m_a$ to get $\bm m_c$. Then, by using $\bm u_{\text{mid}}$, we calculate $\nabla \frac{1}{2}\bm u_{\text{mid}}^2$ on particles. Now, all quantities are on particles,
together with $\bm \Lambda$ and $\bm \Upsilon$ calculated from the previous frame, we get $\bm u_c^*$ following Eq.~\ref{eq:update_equation}.

\paragraph{Coupling with $\bm u_c^*$ and projection} Following the calculation of \(\bm u_c^*\), the coupling method can be performed using this advected divergent velocity field because both solid and fluid are now described using velocity, which is a short-range physical property. Depending on the chosen coupling schemes, either Particle-to-Grid (P2G) or force spreading is performed, as detailed in Section~\ref{sec:time_integration}. After this, coupling force $\bm f_c$ with external forces for a single frame is added to the grid. The velocity on the grid can then be used for Poisson projection to enforce a divergence-free condition.

\paragraph{Summary}
In sum, by accumulating $\bm \Lambda$ and $\bm \Upsilon$ on each particle, we achieve a divergent velocity field from impulse using flow map advection, which substitutes the semi-Lagrangian step in typical fluid solvers without requiring the typical projection operation. 
Now we have (1) solid and fluid using the same variable for representation and (2) forces are correctly handled through maintaining the force buffer $\bm \Upsilon$ on particles, coupling between solid and fluid can fall back to utilizing methods used in typical velocity domain without the presence of a flow map.

\section{Time Integration}
\label{sec:time_integration}

We outline our time integration scheme in Alg.~\ref{alg:general_pipeline}. Our pipeline allows any coupling method to be integrated as long as the computation of Eq.~\ref{eq:sfi_u_continuous} especially the coupling force between solid and fluid can be accumulated to our coupling force buffer $\bm \Upsilon$.
Examples with pseudo code of using this pipeline for coupling under MPM or IBM framework are shown in Appendix~\ref{sec:mpm_app} and Appendix~\ref{sec:ibm_app}.
\begin{enumerate}[leftmargin=*]
    \item \textbf{Reinitialization}. \label{item:reinit}\ After every $n$ steps, quantities carried on fluid particles are reinitialized in the manner adopted from PFM. Specifically, particles are uniformly redistributed, and both $\mathcal{F}^f_{[c, a]}$ and $\mathcal{T}^f_{[a, c]}$ are reset to $\mathcal{I}$, with $\bm m_a$ reset to the initial state using $\bm u_b$. In addition, $\bm{\Lambda}_b$, $\bm{\Upsilon}_b$ are emptied too.
    
    \item \textbf{CFL Condition for Fluid}. $\Delta t$ is computed based on the fluid velocity field and the CFL number. 
    
    \item \textbf{CFL Condition for Solid}. A separate sound CFL condition and velocity CFL condition are used for solid to determine $\Delta t^s$ and the number of solid substeps.
    
    \item \textbf{Midpoint Method}. \label{item:midpoint}\ 
    We utilize a leapfrog-style energy preservation temporal integration scheme to enhance vorticity preservation. Examples of the midpoint method for MPM are provided in Alg.~\ref{alg:midpoint_mpm} and IBM in Alg.~\ref{alg:midpoint_ibm}.
    \paragraph{Fluid} We predict $\bm u_{mid}$ by using $\bm u_b$.
    \paragraph{Solid} We simulate half the number of solid substeps, with initial velocities sampled from $\bm u_{b}$, to synchronize with the fluid state.
    \paragraph{Coupling} We utilize the P2G scheme from MPM or IBM force spreading followed by Poisson solving for incompressibility.

    \item \textbf{Fluid Advection}. \label{item:march_fluid}
    We march $\bm x^f$, $\mathcal{T}_{[a, c]}$ and $\mathcal{F}_{[c, a]}$, according to Eq.~\ref{eq:T_mapping}. 
    
    \item \textbf{Compute $\bm m_c$}. We update $\bm m_c$ with $\bm m_a$ and $\mathcal{T}^f_{[a, c]}$, according to Eq.~\ref{eq:update_equation}.

    \item \textbf{Compute $\nabla u^f_c$}. We compute $\nabla u^f_c$ with $\bm u_{mid}$ using
    \begin{equation}
    \label{eq:compute_grad_u_c}
        \nabla \bm u^f_c = \sum_i \nabla w_{ip} \bm u_{mid},
    \end{equation}
    \revv{where $w_{ip}$ is the quadratic weight function \cite{jiang2016material}.}

    \item \textbf{Solid Marching}.\label{item:solid}
    We simulate the remaining half number of solid substeps with initial velocity sampled from $\bm u_{mid}$ to synchronize with the fluid state.

    \item \textbf{Impulse to Velocity Conversion}. We compute $\bm u_c^{f*}$ with $\bm{\Lambda}_b$, $\bm{\Upsilon}_b$, $\bm{u}_{\text{mid}}$ and $\bm m_c$, according to Eq.~\ref{eq:update_equation}.

    \item \textbf{P2G and Coupling}. Compute $\bm u_c^*$ by a P2G process using quantities carried on particles. When using MPM, we perform the P2G process for both fluid and solid using $\bm u_c^f$ and $\bm u_c^n$ (details provided in Section~\ref{sec:mpm} and Appendix~\ref{app:mpm_sup}) on fluid particles and $\bm u^s_c$ on solid particles. When using IBM, force is distributed to the velocity field after the P2G process. 

    \item \textbf{Add External and Coupling Forces}. Coupling forces $\bm f_c$ and external forces like gravity $\rho \bm g$ and viscosity $\upsilon \Delta \bm{m}_c$ are added to the grid.

    \item \textbf{Poisson Solving}. We solve Poisson equation to ensure divergence-free condition.
    
    \item \textbf{Buffer Update}. We update $\bm \Upsilon_c$ and $\bm \Lambda_c$, according to Eq.~\ref{eq:update_equation}.
\end{enumerate}

\begin{algorithm}[t]
\caption{Time Integration}
\label{alg:general_pipeline}
\begin{algorithmic}[1]
\For{$k$ in total steps}
\State \revv{Reinitialization every $n$ steps}; \hfill 

\State Compute $\Delta t$ with $\bm{u}_b$ and the CFL number;
\State Determine $\Delta t^s$ and number of substeps for solid;

\State Estimate midpoint velocity $\bm u_{\text{mid}}$;
\hfill $\triangleright$ Alg.~\ref{alg:midpoint_mpm}/\ref{alg:midpoint_ibm}

\State March $\bm x^f$, $\mathcal{T}^f_{[a, c]}$ $\mathcal{F}^f_{[c, a]}$ with $\bm{u}_{\text{mid}}$ and $\Delta t$; \hfill $\triangleright$ Eq.~\ref{eq:T_mapping}
\State Compute $\bm m_c$ with $\bm m_a$ and $\mathcal{T}^f_{[a, c]}$; \hfill $\triangleright$ Eq.~\ref{eq:update_equation}
\State Compute $\nabla \bm u^f_c$ using $\bm{u}_{\text{mid}}$; \hfill $\triangleright$ Eq.~\ref{eq:compute_grad_u_c}

\State March solid based on solid simulation methods;


\State  Compute $\bm u_c^{f*}$ with $\bm{\Lambda}_b$, $\bm{\Upsilon}_b$, $\bm{u}_{\text{mid}}$, $\bm m_c$; \hfill $\triangleright$ Eq.~\ref{eq:update_equation} 

\State Compute $\bm u_c^*$ by transferring $\bm u_c^{f*}$(, $\bm u_c^n$, $\bm u_c^s$) and $\nabla \bm u^f_c$ to grid; \\ \hfill $\triangleright$ Eq.~\ref{eq:P2G}
\State Add single step coupling forces $\bm f_c$ and external forces;

\State $\bm u_c, p_c \gets \textbf{Poisson}(\bm u_c^*)$;

\State Update $\bm \Upsilon_c$ and $\bm \Lambda_c$ with $\bm u_c$, $\mathcal{F}^f_{[c, a]}$, $\bm f_c$ and $p_c$; \hfill $\triangleright$ Eq.~\ref{eq:update_equation}

\EndFor{}
\end{algorithmic}
\end{algorithm}

\begin{figure*}[t]
\centering
\begin{minipage}{.49\linewidth}
  \includegraphics[width=\linewidth]{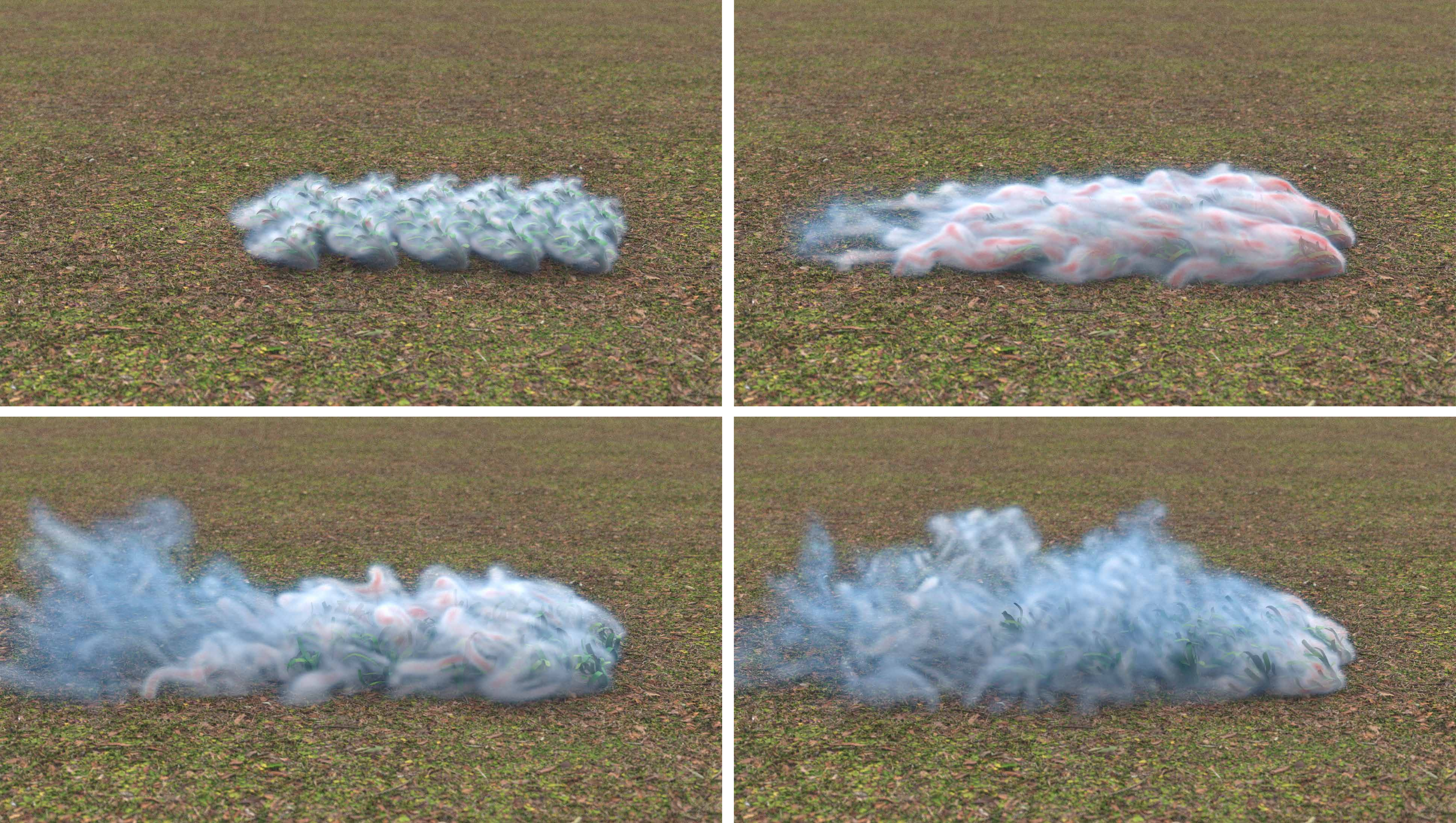}
  \caption{Under incoming flow with variant velocity, turbulent surface flow is created from the grass geometry. Vorticity is being visualized here for illustration of the turbulent flow.}
 \label{fig:3d_grass}
\end{minipage}
\hspace{.01\linewidth}
\begin{minipage}{.49\linewidth}
  \includegraphics[width=\linewidth]{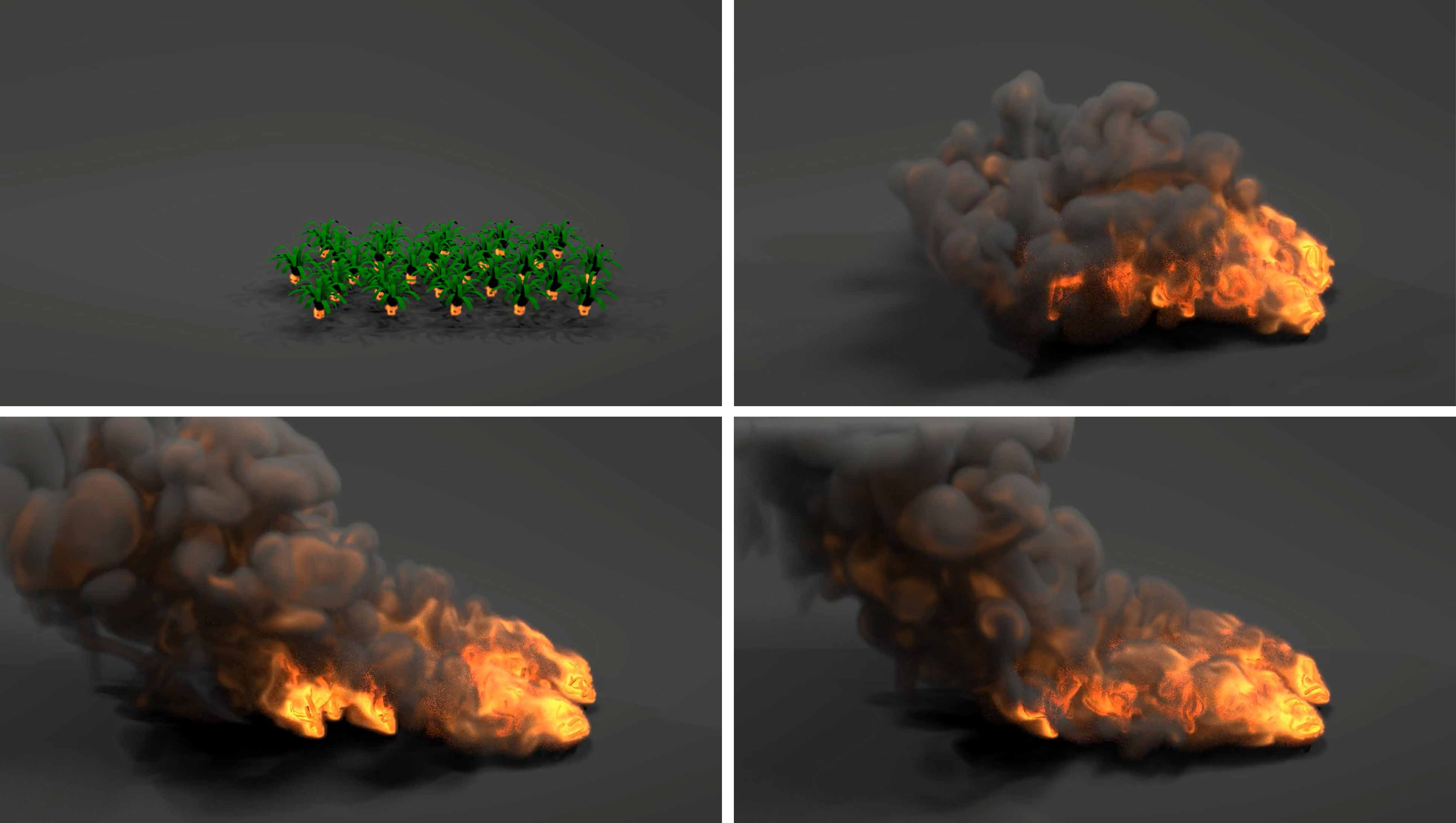}
 \caption{\revv{We show the fire ignition on grassland leading to intense smoke and fire production. We see interesting vortical structures formed within the visualized smoke and fire regions.}}
 \label{fig:3d_grass_fire}
\end{minipage}
\end{figure*}



\section{Coupling Framework Examples}
This section demonstrates two coupling frameworks on MPM and IBM using our flow-map coupling method. Each framework calculates the coupling forces that can be exchanged between solid and fluid particles (implicitly for MPM and explicitly for IBM) and then accumulates these forces on our particle force buffers.

\subsection{Coupling with MPM}
\label{sec:mpm}
This section demonstrates how to adapt our coupling method to the MPM framework. Couplings within the MPM framework are realized through P2G and G2P procedures and pressure projection for momentum exchange and coupling force calculation. Because the coupling force for MPM implicitly inherits from the pressure gradient in the projection step, our accumulation for $\bm \Lambda$ provides the buffer for accumulating coupling force under the MPM framework in flow map settings. Using our scheme provided in Section~\ref{sec:time_integration}, fluid particles carrying impulse variables described under a long-range flow map can be converted to a short-range velocity in the same physical model as MPM particles. With this unified presentation, we perform P2G using typical P2G procedures for solid particles $\bm u_c^s$ and fluid particles $\bm u_c^f$. For simplicity, we use $\bm u_c^p$ to represent velocity for both kinds of particles:
\begin{equation}
\label{eq:P2G}
    \bm u_i \gets \sum_p w_{ip}(\bm u_c^p + \nabla \bm u_c^p (\bm x_i - \bm x_p)) \,/\, \sum_p w_{ip}, 
\end{equation}
where weight kernel is chosen to be quadratic kernel in \cite{jiang2016material} and $\bm x_i$ denotes MAC grid face center location.

The G2P process is performed for fluid and solid particles in the advection step. One difference from classical MPM simulation is that we perform RK4 time integration for advection for better energy conservation. The G2P process is performed with 
\begin{equation}
\label{eq:reinit_imp_long}
    \bm u^p \gets \sum_i w_{ip} \bm u_i.
\end{equation}

\subsection{Coupling with IBM}
\label{sec:ibm}
The IBM coupling happens between solid and fluid through the exchange of elastic force spread through a smoothed Dirac-Delta kernel $\delta_h(\bm x)=\frac{1}{h^3}\phi(\frac{x}{h})\phi(\frac{y}{h})\phi(\frac{z}{h})$, where $\phi(r)$ is defined as:
\begin{equation}
\phi(r)=\left\{\begin{array}{ll}
\frac{1}{4}\left(1+\cos \left(\frac{\pi r}{2}\right)\right) & |r| \leqslant 2 \\
0 & \text { otherwise }
\end{array},\right.
\end{equation}
Due to this explicit definition of coupling force, we can easily integrate the coupling force using our particle buffer $\bm \Upsilon$ at each step. After we convert from the long-range mapped impulse variable to the short-range velocity for fluid particles and perform P2G for fluid particles, solid coupling forces are spread to the grid using the smoothed Dirac-Delta kernel as:
\begin{equation}
    \bm f^{\text{grid}} = \int \bm f^{\text{solid}} \delta(\bm x^{\text{grid}} - \bm X^{\text{solid}})
\end{equation}
This force is accumulated to $\bm \Upsilon$ at the end of this time step. The divergence-free velocity field interpolated to solid particles through the kernel function is used to march the solid particles following:
\begin{equation}
    \bm u^{\text{solid}} = \int \bm u^{\text{grid}} \delta(\bm x^{\text{grid}} - \bm X^{\text{solid}})
\end{equation}
Here, $\bm f^{\text{grid}}$ corresponds to the force that is added to the velocity field after the impulse-to-velocity conversion, and $\bm u^{\text{grid}}$ represents the divergence-free velocity field after Poisson projection.

\revv{For implementation details, all advections are performed using the 4th order of the Runge-Kutta (RK4) method in \cite{deng2023fluid,zhou2024eulerianlagrangian} to track flow maps on particles. Specifically, we follow the common practice in APIC that uses the gradient of kernel function to calculate $\nabla \bm u$.} We provide more details for implementing MPM and IBM in Appendix~\ref{app:mpm_sup} and Appendix~\ref{app:ibm_sup}.


\begin{figure*}[t]
\centering
\begin{minipage}{.49\linewidth}
  \includegraphics[width=\linewidth]{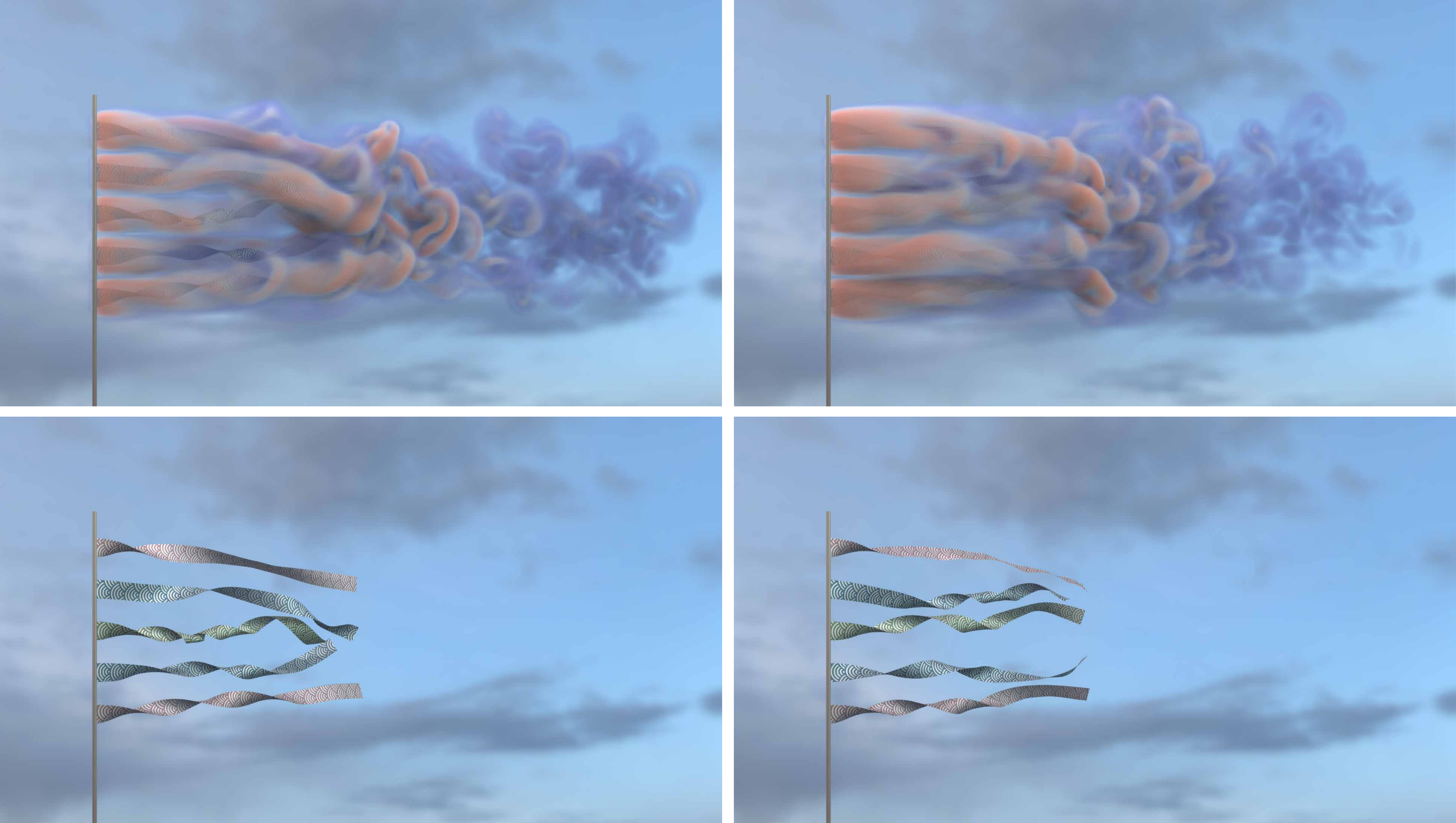}
  \caption{In this experiment, we show a long silk flag affected by an incoming flow with variant velocity. Interesting spiral-shaped vortices are formed. We refer readers to our supplementary video for the dynamic motion of the vortices.}
 \label{fig:3d_silk}
\end{minipage}
\hspace{.01\linewidth}
\begin{minipage}{.49\linewidth}
  \includegraphics[width=\linewidth]{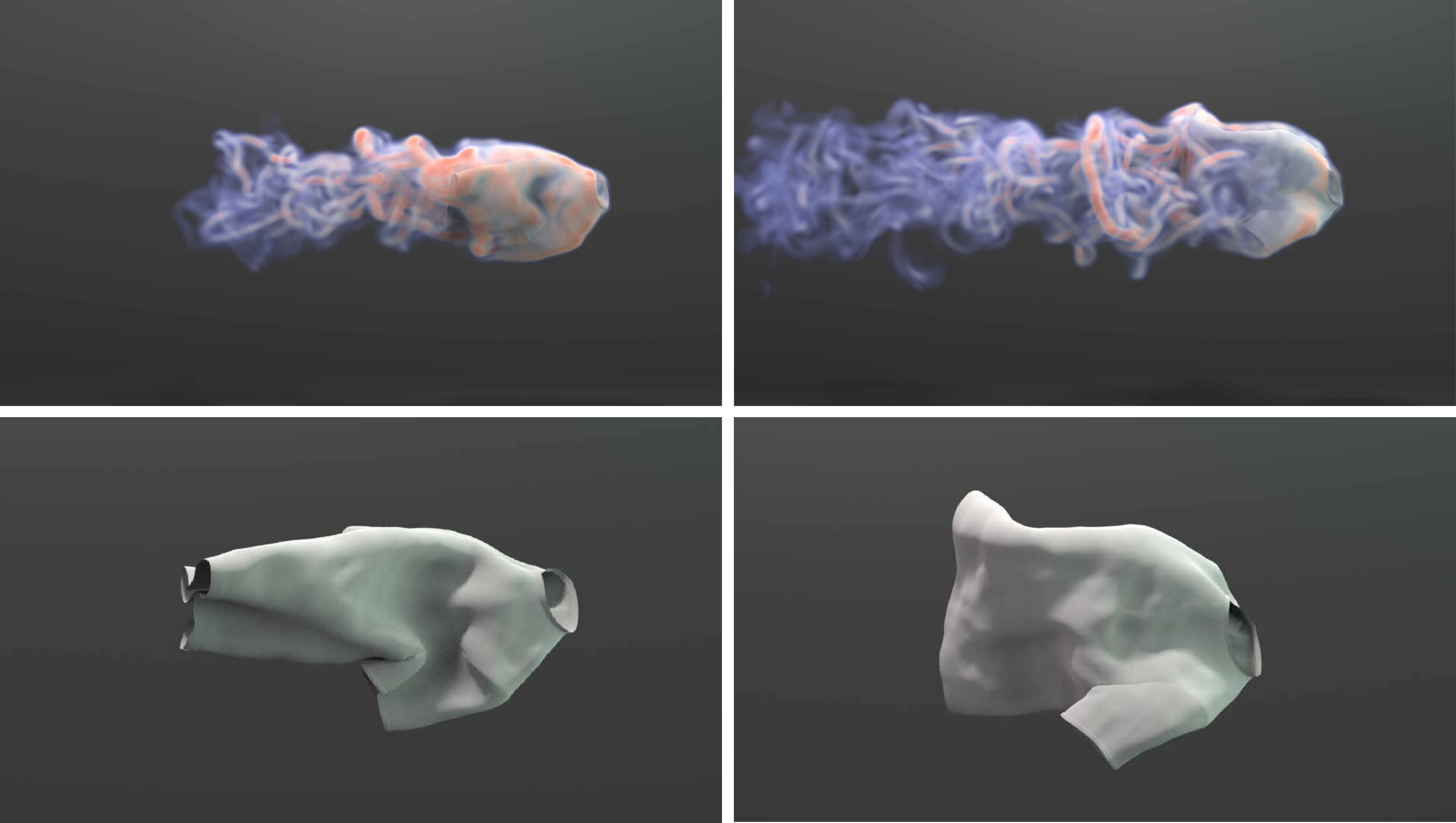}
 \caption{\revv{T-shirt in wind field: we show high-frequency oscillation on the T-shirt geometry can be clearly observed. Such wrinkles of the T-shirt geometry are driven by the vorticity shredding on the surface which is automatically formed due to our coupling mechanism.}}
 \label{fig:3D_shirt}
\end{minipage}
\end{figure*}

\begin{figure}[t]
 \centering
 \includegraphics[width=.99\columnwidth]{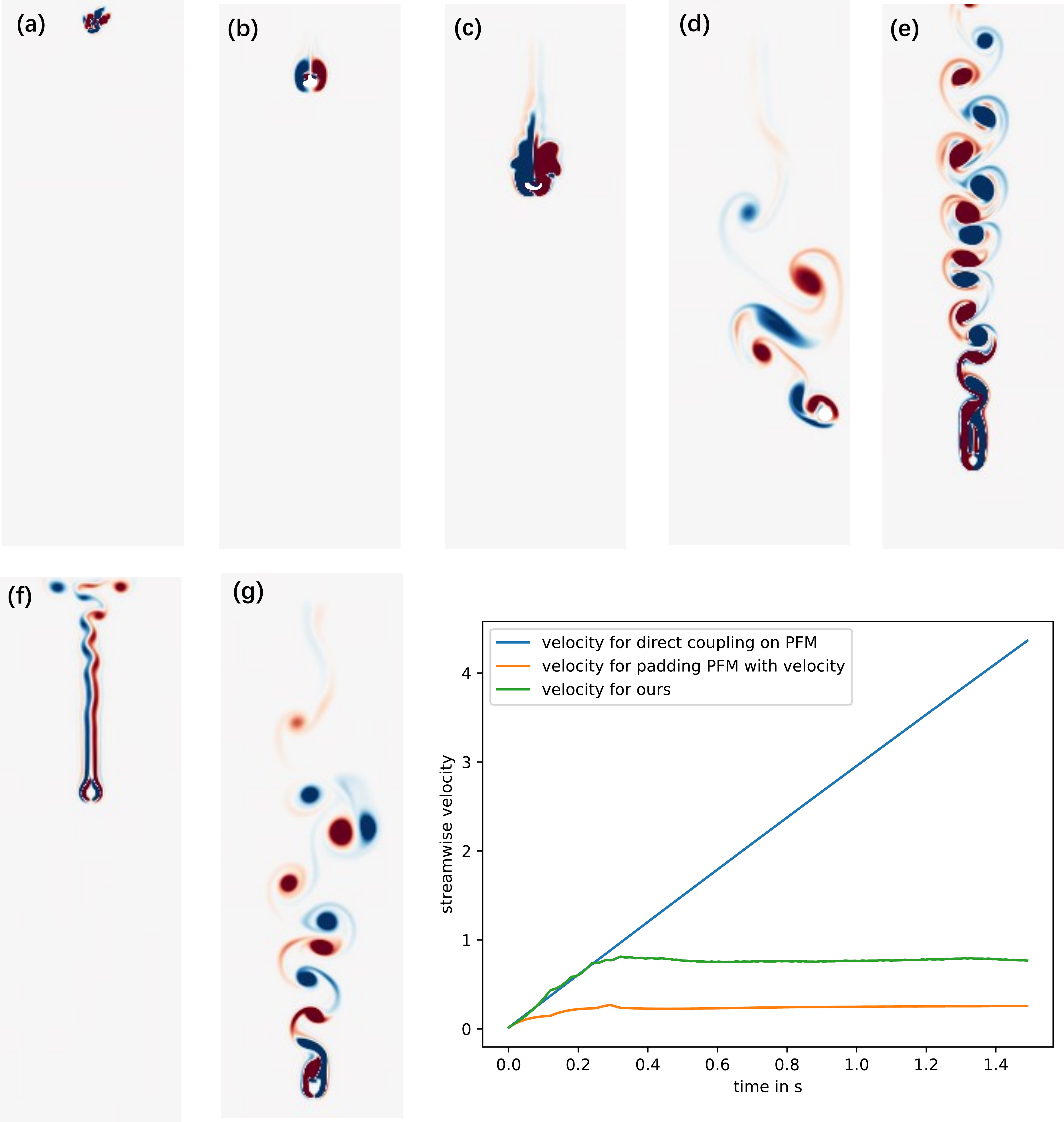}
 \caption{On the left figure, we illustrate several approaches: (a) Direct GFMC, (b) Single-Step Heaviside GFMC,(c) Accumulative Heaviside GFMC, (d) Tuned Accumulative Heaviside GFMC, (e) Direct HFMC, (f) Padded Velocity HFMC, and (g) our method. In the right figure, we demonstrate that direct hybrid flow map coupling does not converge to a velocity, which is expected due to the fluid-to-solid forces.}
 \label{fig:diff_method}
\end{figure}


\begin{figure*}[t]
 \centering
 \includegraphics[width=.99\textwidth]{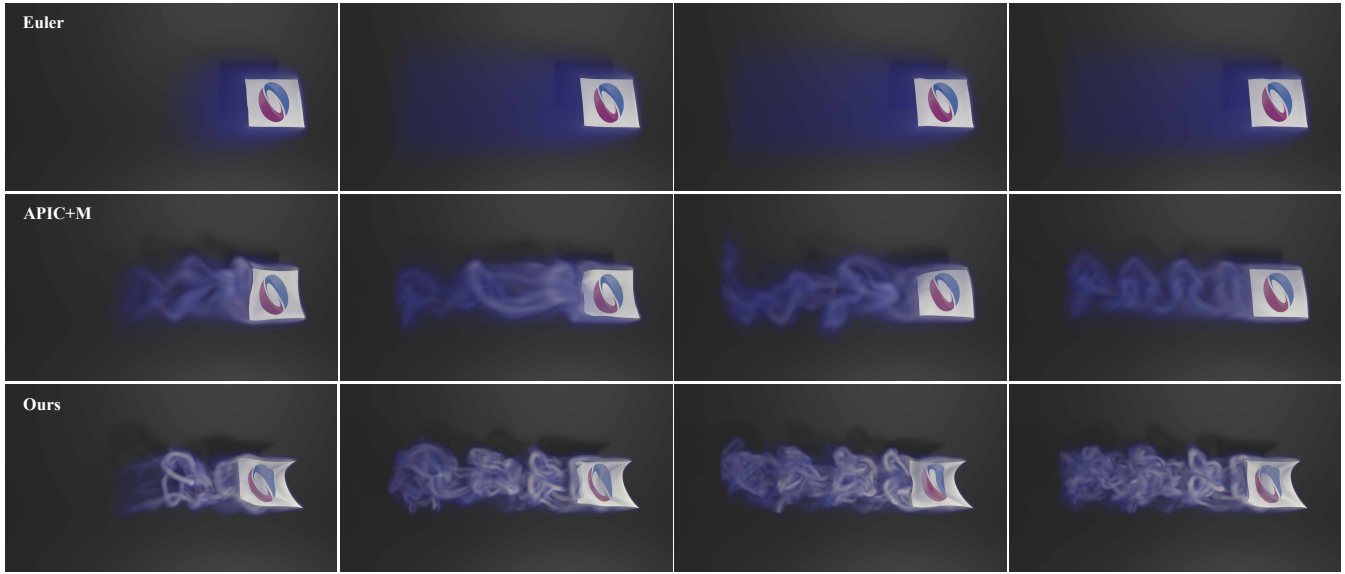}
 \caption{We compare our method, Euler's method, and APIC+M. We observed no vorticities in Euler. APIC+M forms blurred vortices, and ours shows the clearest structure of vortices.}
 \label{fig:3D_cloth_compare}
\end{figure*}

\begin{figure*}
\centering
\begin{minipage}{.49\linewidth}
  \includegraphics[width=\linewidth]{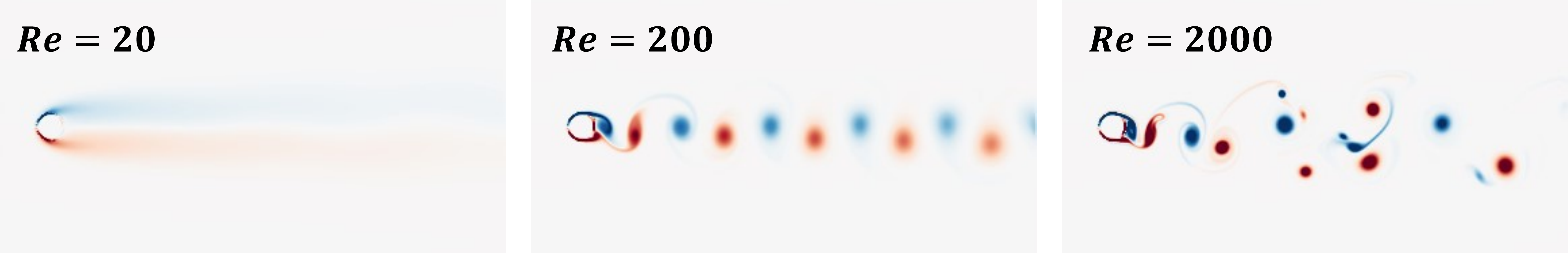}
 \caption{Karman Vortex street under different Re number which is calculated from viscosity $\mu = 4\times10^{-4}$, $4\times10^{-5}$, $4\times10^{-6}$ with geometry radius being 0.05 and dominate velocity being 0.16.}
 \label{fig:diff_re}
\end{minipage}
\hspace{.01\linewidth}
\begin{minipage}{.49\linewidth}
  \includegraphics[width=\linewidth]{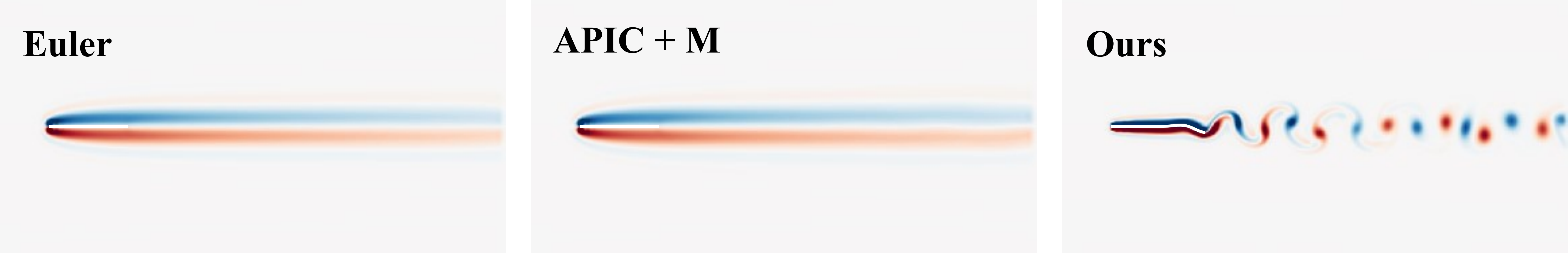}
 \caption{This experiment compares our method with traditional fluid models within the IBM framework. Using a 2D mass-spring system with an incoming velocity of 0.16, our method demonstrates the formation of a vorticity street.}
 \label{fig:2D_ibm_compare}
\end{minipage}
\end{figure*}
\section{Ablation Studies}
\label{sec:ablation}


In this experiment, we implemented six different alternatives in addition to our proposed method, including coupling on grid-based flow map methods with/without RMT \cite{rycroft2020reference} and force accumulation, hybrid flow map methods directly using P2G and with a thin layer of velocity particles. As shown in Figure~\ref{fig:diff_method}, we tested these seven implementations in the example of a falling sphere immersed in the fluid and analyzed the results of their simulations. We use a simulation domain size of 128x384 and set gravity to 3. We use 0.03 as the radius for the sphere with a density ratio of 15:1 and $E=5\times 10^3$ and $\mu = 0.3$ throughout all the tests.

For grid-based flow map methods, we implemented the reference map method described in \cite{rycroft2020reference}, which includes flow map extrapolation and Heaviside blending for (1) fluid/solid density and (2) fluid stress $\tau^f$ and solid $\tau^s$ on grid-based flow map. At each step, a level set is created using MPM particles, and a Heaviside function ($H$) is calculated based on this level set on the grid. This Heaviside function is utilized to blend MPM solid stress $\tau^s$ with fluid viscosity stress $\tau^f$, which is defined as $\mu^{f} (\nabla \bm u + (\nabla \bm u)^T)$, and to blend solid density $\rho^{s}$ with fluid density $\rho^{f}$. The blended value is defined as:
\begin{equation}
    \begin{aligned}
        \tau &= \tau^f + H_{\varepsilon}(\tau^s - \tau^f)\\
        \rho &= \rho^{f} + H_{\varepsilon}(\rho^{s} - \rho^{f})\\
    \end{aligned}
\end{equation}
and 
\begin{equation}
    H_{\varepsilon}(\phi)= \begin{cases}0 & \text { if } \phi \leqslant-\varepsilon, \\ \frac{1}{2}\left(1+\frac{\phi}{\varepsilon}+\frac{1}{\pi} \sin \frac{\pi \phi}{\varepsilon}\right) & \text { if }|\phi|<\varepsilon, \\ 1 & \text { if } \phi \geqslant \varepsilon,\end{cases}
\end{equation}
where $\phi$ is the level set value and $\varepsilon$ denotes the width of the blurring zone and is typically set at 2.5$\Delta x$, as noted in \cite{rycroft2020reference}. The blurred $\tau$ is then used for solid elastic force calculation and $\rho$ for P2G calculation. To utilize this blending scheme, specifically to calculate $\tau^s$ outside solid, we sample a narrowband of blur-zone particles in MPM simulation, and sampling follows the same way as described in Section~\ref{sec:resample}. Complicatedly, using this blending requires extrapolation of $\mathcal{F}^{\text{particle}}$ in the blur zone to calculate $\tau^s$ outside solid. This task lacks clear physical meaning, and the way we do this is to directly copy the $\mathcal{F}$ from the nearest surface particle to the particle in the blur zone.  Forces calculated from P2G using blurred $\mathcal{F}$ are applied to impulse field advected by flow map. 


\paragraph{(a) Direct Grid Flow Map Coupling (Direct GFMC)} We directly use the deformation gradient on solids to calculate force, without blending with the forward map gradient $\mathcal{F}$ on fluids stored on the grid. We show in Figure~\ref{fig:diff_method}(a) that numerical error dominates.
\paragraph{(b) Single-Step Heaviside GFMC} We add force for just a single frame within the grid-based flow map framework using the RMT method. We show in Figure~\ref{fig:diff_method}(b) that such a method results in a slower falling speed than expected but also causes the solid to behave much softer than it should.
\paragraph{(c) Accumulative Heaviside GFMC} We add the accumulated force without mapping the force to the first frame within the flow map framework together with the RMT method. In Figure~\ref{fig:diff_method}(c), we illustrate that this leads to numerical instability
\paragraph{(d) Tuned Accumulative Heaviside GFMC} We use the same method as (c) but adjust the parameters to use smaller timestep, density ratio and gravity. However, distortions in flow map extrapolation become predominant, culminating in incorrect behavior as depicted in Figure~\ref{fig:diff_method}(d).
\paragraph{(e) Direct Hybrid Flow Map Coupling (Direct HFMC)} We use the impulse for fluid particles and directly P2G with solid particles carrying velocity and solve Poisson together. This method does not result in velocity convergence, as seen in Figure~\ref{fig:diff_method}(e) and the right figure of Figure~\ref{fig:diff_method}. This issue arises due to the differing representations of fluid and solid, specifically impulse versus velocity.
\paragraph{(f) Padded Velocity HFMC} We pad solid particles with a layer of fluid particles carrying velocity instead of impulse but didn't deal with fluid particles distant from the solid. Velocity convergence is observed. However, this approach makes fluid behavior anomalous due to the different models used for describing fluid dynamics as observed in Figure~\ref{fig:diff_method}(f).

\paragraph{(g) Ours} We show our method correctly couples fluid and MPM solid without the issues presented above in Figure~\ref{fig:diff_method}(g).



Based on the observations above, to correctly couple solid and fluid under the flow map framework, a method must ensure that (1) fluid and solid have the same physical representation when performing coupling and (2) forces are correctly managed within the flow map framework. Our framework successfully satisfies these requirements.

\section{Validation and Comparison}
In this section, we first verify the correctness of our external force treatment by validating it through the Karman Vortex Street experiment, which is run under different viscosity parameters leading to various Reynolds numbers and then on a cylinder with different density sedimentation that converges to different velocities. We then compare our method against Euler's method and affine particle-in-cell (APIC) as shown in Fig.~\ref{fig:3D_cloth_compare}. As highlighted in \cite{nabizadeh2022covector,deng2023fluid}, a symmetric energy conservation scheme like the Leapfrog/Verlet method is crucial for vorticity conservation. We enhance the APIC method by adding a midpoint prediction, denoted as APIC+M, for a fair comparison.

\subsection{2D Viscosity Test}
We demonstrate that we can correctly handle viscosity with our treatment for external force, as discussed in Section~\ref{sec:impulse_to_vel_numerical}. In Figure~\ref{fig:diff_re}, we present the Karman Vortex Street phenomenon under different Reynolds numbers, specifically 20, 200, and 2000.


\begin{figure*}[t]
\centering
\begin{minipage}{.49\linewidth}
  \includegraphics[width=\linewidth]{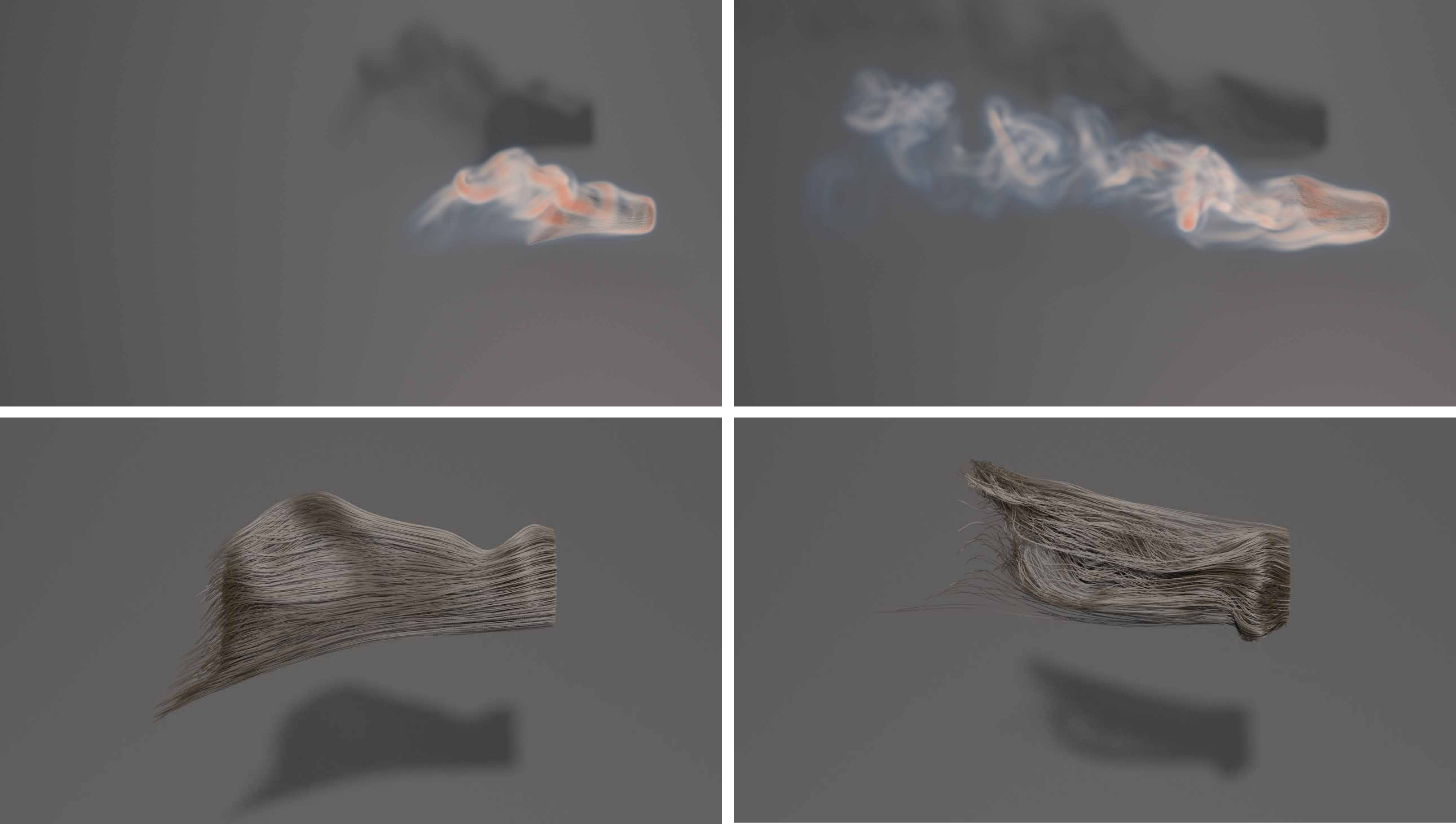}
  \caption{We simulate hair using the model described in \cite{bender2014survey}, employing XPBD and coupling it with our particle flow map fluid solver. Complex vortex structures emerge under the influence of a constant incoming flow, causing the hair to be lifted due to the flow's speed.}
 \label{fig:3d_hair}
\end{minipage}
\hspace{.01\linewidth}
\begin{minipage}{.49\linewidth}
  \includegraphics[width=\linewidth]{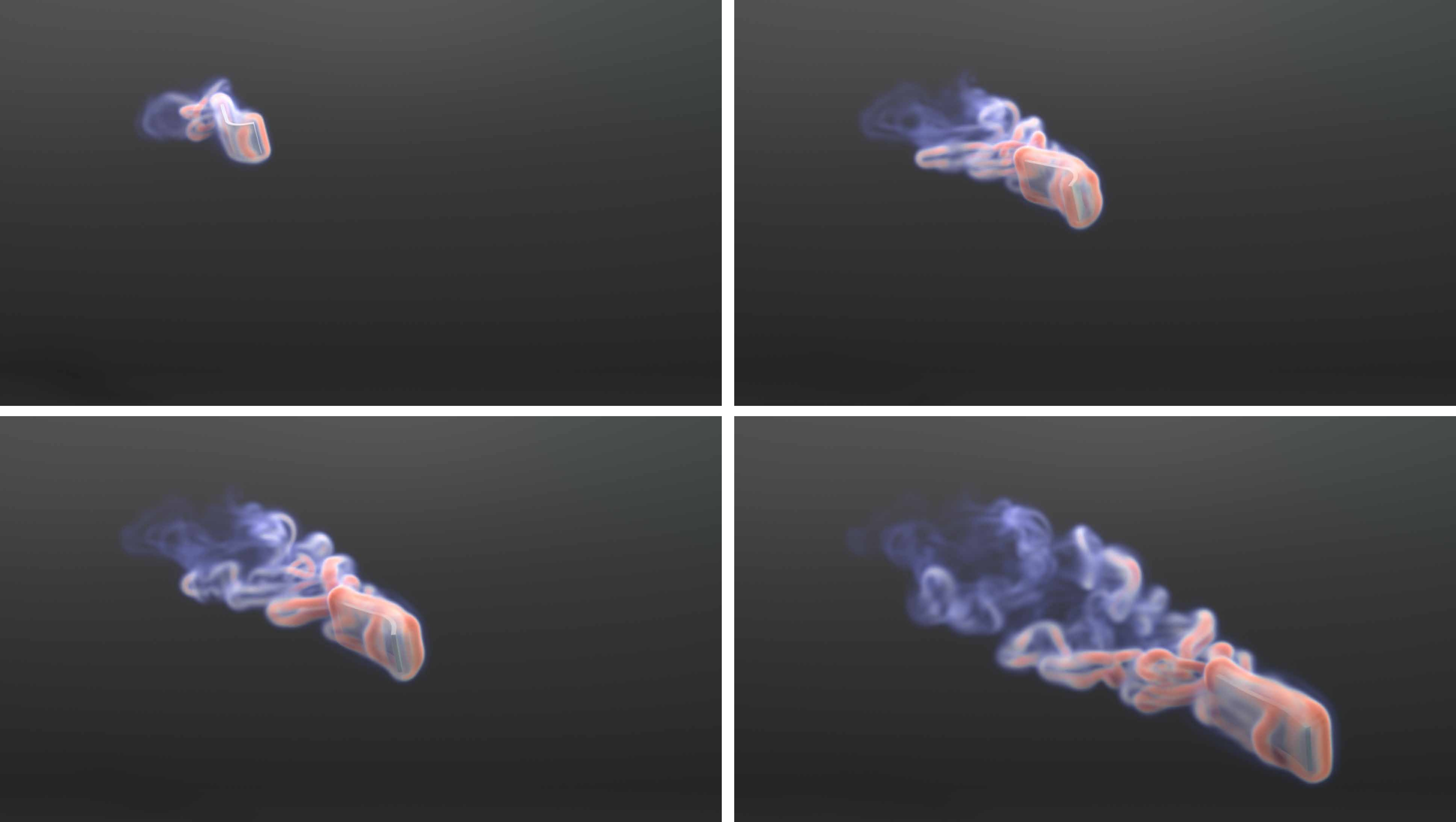}
 \caption{We show the result of our method combining the active strain method to generate a self-contraction soft body simulated by MPM, creating a fish-like movement to move forward. Vortices behind the fishtail form a similar pattern as observed in \cite{lin2019fluid}.}
 \label{fig:3d_fish}
\end{minipage}
\end{figure*}

\begin{figure*}
\centering
\begin{minipage}{.49\linewidth}
  \includegraphics[width=\linewidth]{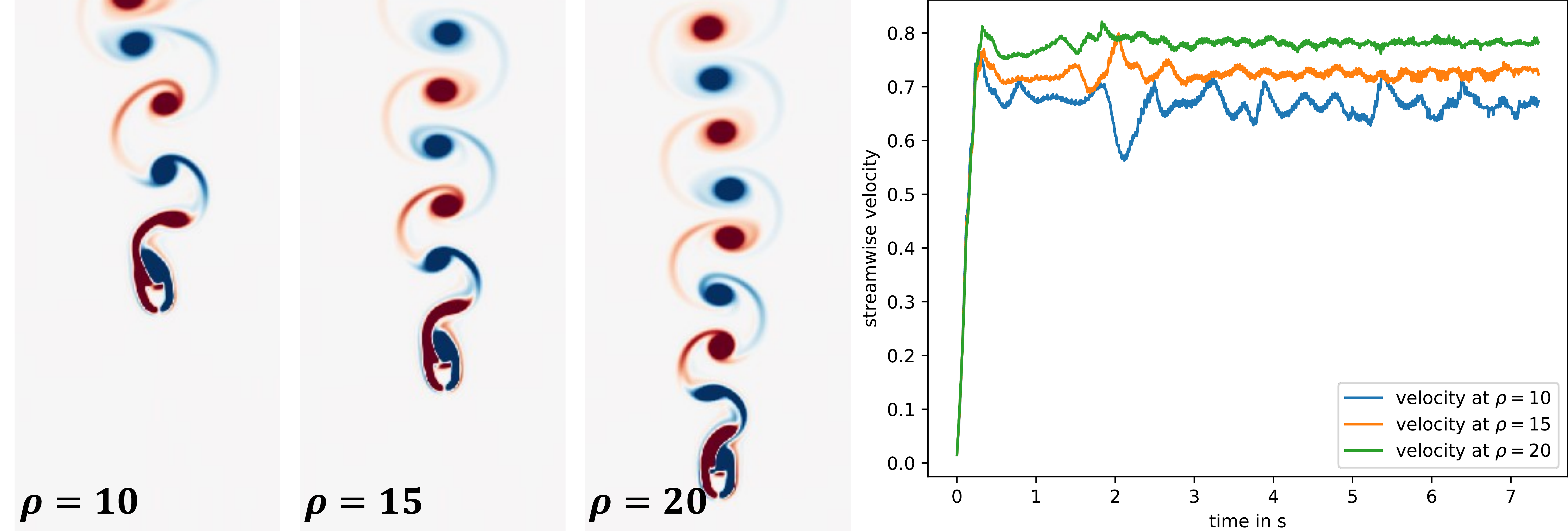}
 \caption{The image on the left illustrates a comparison of the cylinder positions at time $t=7.5$, highlighting the effects of different density settings on their behavior. All other parameters remain constant except for the density. The right image shows the convergences of the terminal velocity of the cylinders.}
 \label{fig:diff_rho}
\end{minipage}
\hspace{.01\linewidth}
\begin{minipage}{.49\linewidth}
  \includegraphics[width=\linewidth]{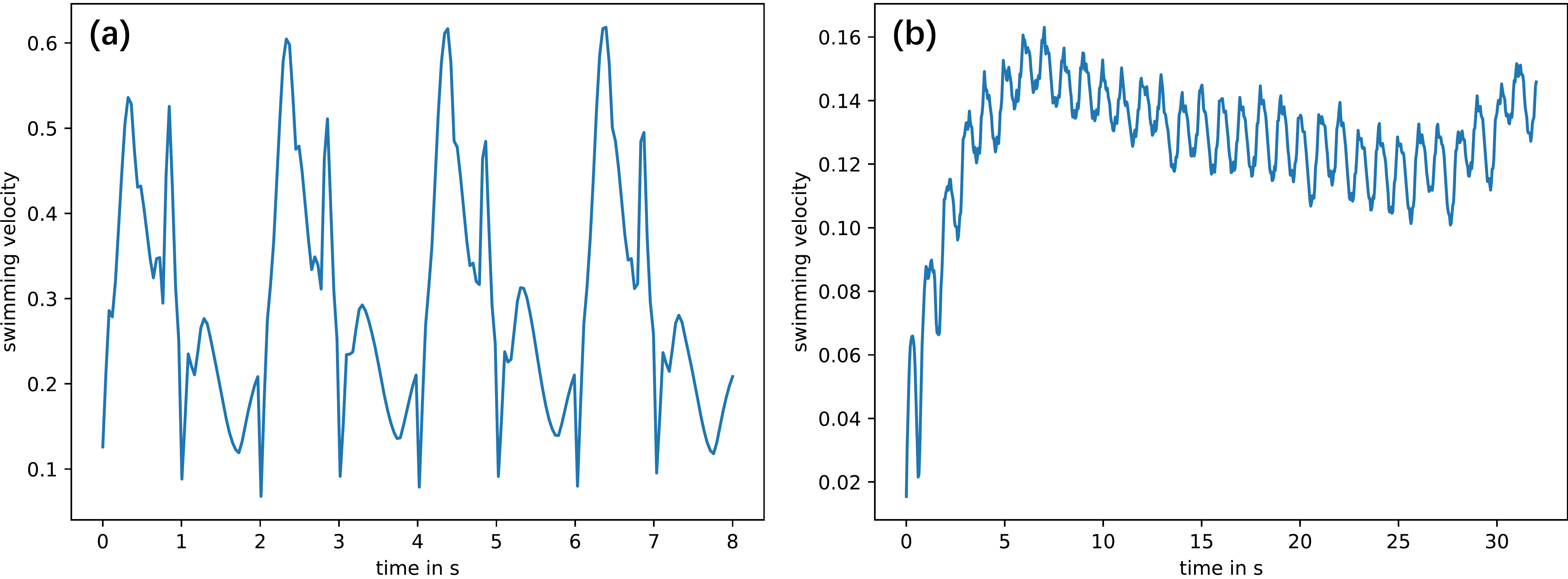}
 \caption{This figure shows the solid velocity of our flapping swimmer and fish model shown in 2D swimmer experiment in Sec.~\ref{sec:swim_result} and fish experiment in Sec.~\ref{sec:fish_result}. The velocity of this plot is calculated with finite difference using positions of two consecutive frames.}
 \label{fig:2D_fish_swimmer_vel}
\end{minipage}
\end{figure*}

\WrapFig{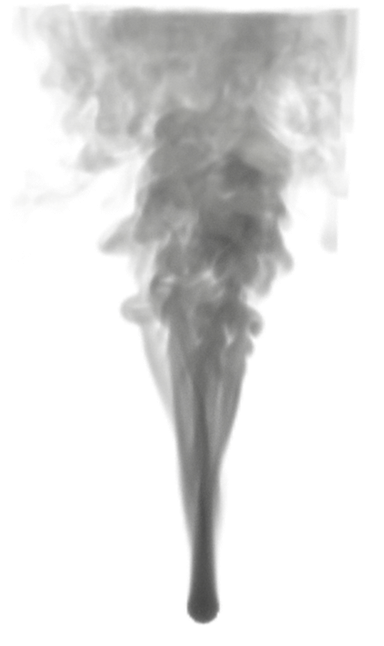}{\revv{Turbulent smoke plume driven by buoyancy.}}{fig:3d_bouyancy}{.15}{1.4}

\subsection{3D Bouyancy Test}
\revv{
We demonstrate that our algorithm can correctly handle external forces such as buoyancy in a 3D plume example. As shown in Figure~\ref{fig:3d_bouyancy}, a turbulent smoke plume rises due to buoyancy and produces complex flow details.
}

\subsection{2D Cylinder Sediment}
We demonstrate that with proper density treatment in handling impulse-to-velocity conversion and Poisson projection with Eq.~\ref{eq:density}, cylinders with different densities will converge to different velocities. The converged velocity is proportional to the cylinder density, as shown in Figure~\ref{fig:diff_rho}.

\subsection{2D Swimmer}
We compare our method against Euler's and APIC+M under the setting of a periodic flapping swimmer. We first introduce the method we used to create such motion, which is utilized not only for this comparison but also for 2D and 3D fish experiments.

\begin{figure*}[t]
\centering
\begin{minipage}{.49\linewidth}
  \includegraphics[width=\linewidth]{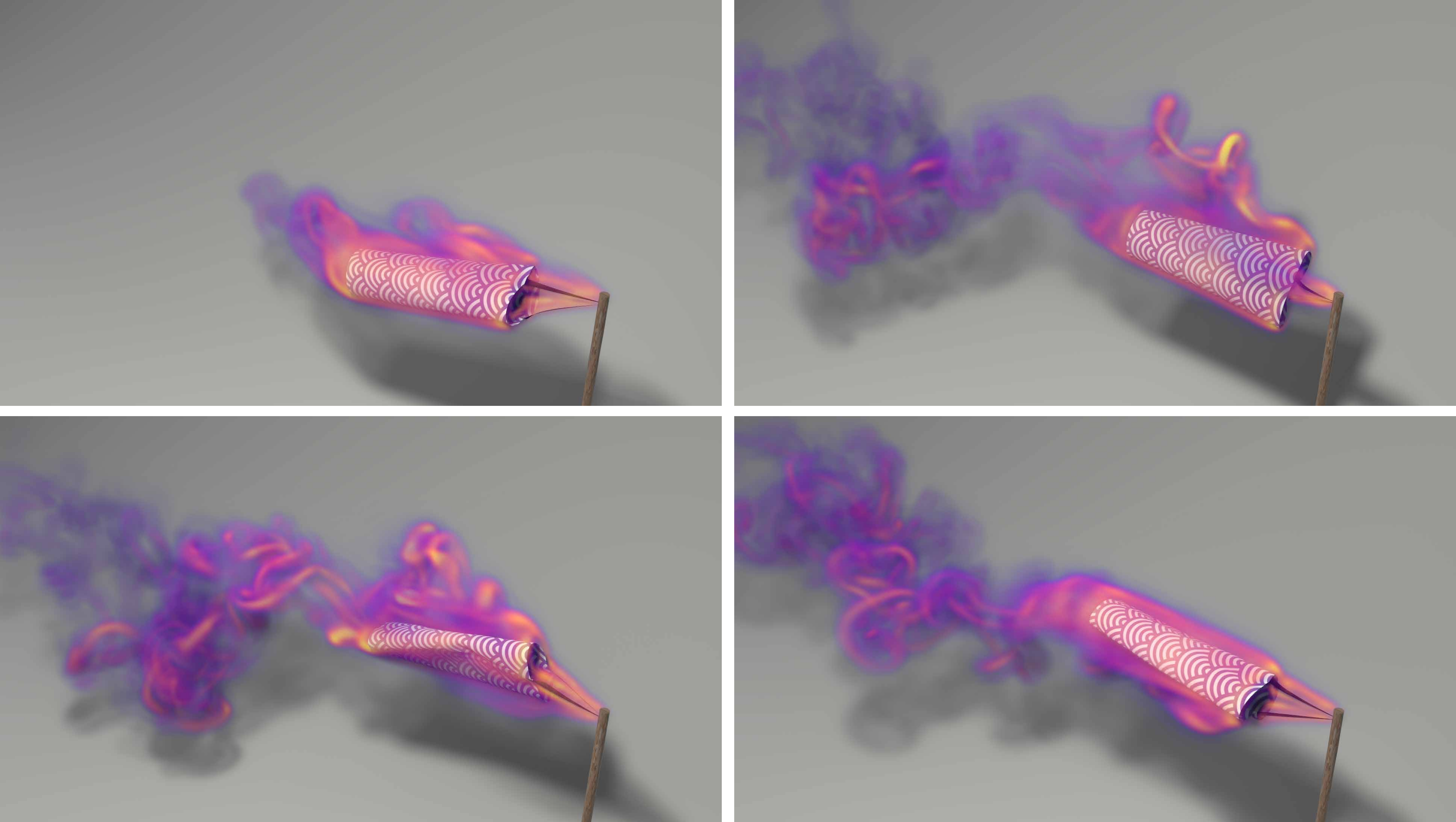}
  \caption{We simulate a Koinobori, represented by a cylindrical shape adorned with a traditional Japanese cloud pattern, being lifted by a constant incoming flow. The waving motion of the Koinobori generates intricate vortex structures along its surface and tail, with the vorticity being visualized to highlight the turbulent flow dynamics.}
 \label{fig:3d_cyn}
\end{minipage}
\hspace{.01\linewidth}
\begin{minipage}{.49\linewidth}
  \includegraphics[width=\linewidth]{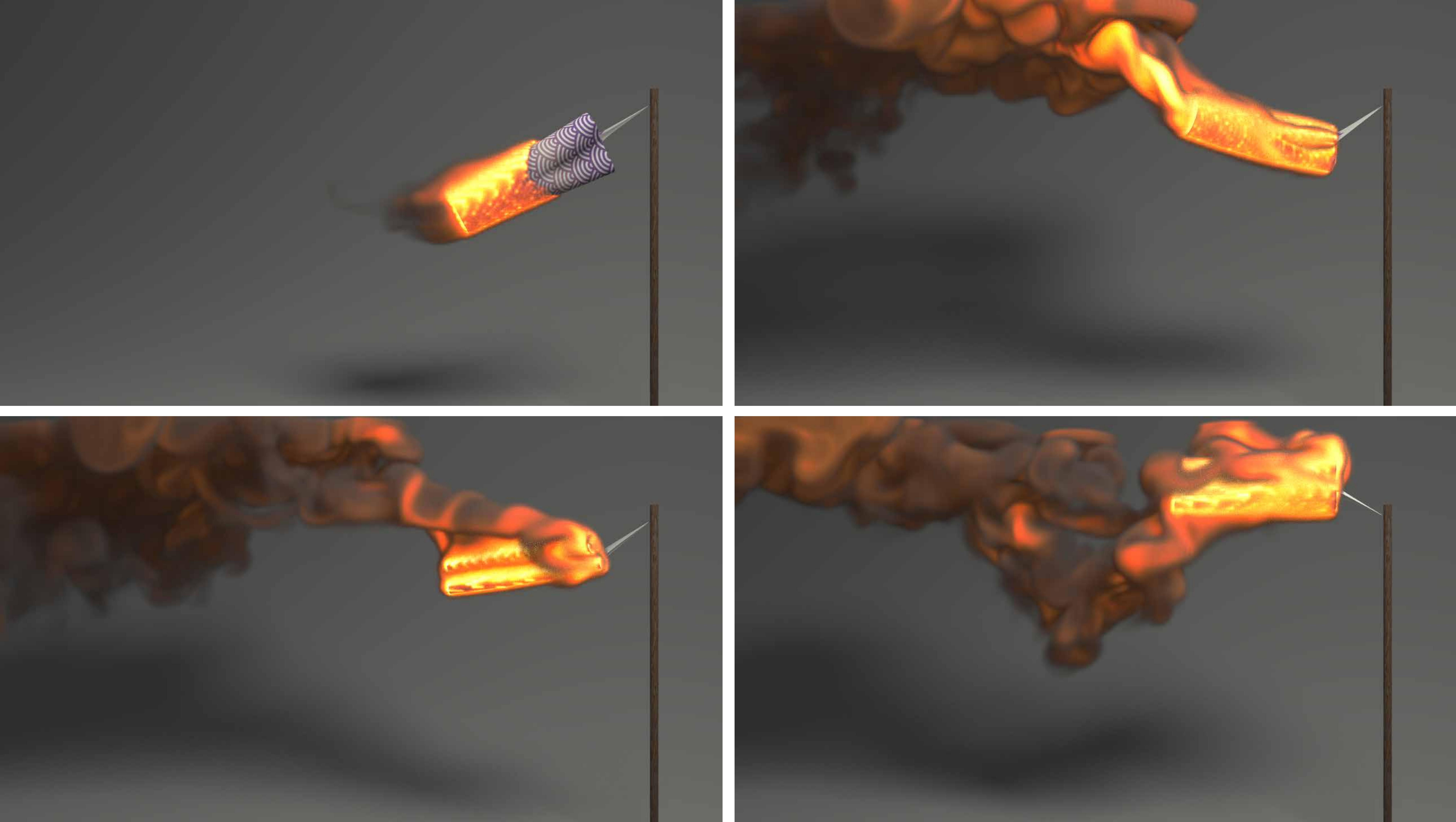}
 \caption{In this experiment, we incorporate combustion into the Koinobori simulation, leading to the vivid visualization of fire and smoke colored with black-body radiation. As the flag waves under fiery conditions, it generates intricate vortex patterns and dynamic interactions between the flames, smoke, and the flag itself.}
 \label{fig:3d_fire}
\end{minipage}
\end{figure*}

\paragraph{Active strain contraction}
\label{sec:active_strain}
We show our method adopting the active-strain method \cite{lin2019fluid} to create periodic self-contraction, which moves solid through interaction with the fluid. The active strain approach imposes contractive strains to drive elastic deformation following a multiplicative decomposition of the deformation gradient tensor. This method defines a $\mathcal,{F}_a$ as an active deformation tensor. To apply actuation on deformation gradient $\mathcal{F}$ is by assuming $\mathcal{F}_{\text{total}} = \mathcal{F} \cdot \mathcal{F}_a$ following multiplicative decomposition \cite{lee1967finite, lee1969elastic}. Therefore the deformation gradient $\mathcal{F}$ used in stress calculation is given by:
\begin{equation}
    \mathcal{F} = \mathcal{F}_{\text{total}} \cdot (\mathcal{F}_a)^{-1},
\end{equation}
where $\mathcal{F}_{\text{total}}$ is the deformation gradient advected from previous timestep and $\mathcal{F}_a$ is the actuation applied.

We use the force model in \cite{lin2019fluid} for $\mathcal{F}_a$. By assuming principle contraction in first dimension, $\mathcal{F}_a$ is defined as $\mathcal{F}_a = \text{diag}[\lambda, \lambda^{-1}]$ in 2D and $\mathcal{F}_a = \text{diag}[\lambda, \sqrt{\lambda^{-1}}, \sqrt{\lambda^{-1}}]$ in 3D where $\lambda < 1$. Contraction in other principle axes can also be applied similarly.

In calculation of $\lambda$, we use:
\begin{equation}
\lambda= \begin{cases}1-\alpha \sin \left(\frac{2 \pi t}{T}\right) \exp \left(-\frac{h-y}{d_0}\right), & 0 \leq \mathrm{t} \leq \mathrm{T} / 2, \\ 
1-\alpha \sin \left(\frac{2 \pi t}{T}\right) \exp \left(-\frac{y}{d_0}\right), & \mathrm{T} / 2<\mathrm{t} \leq \mathrm{T},\end{cases}
\end{equation}
for flapping swimmer experiment and use:
\begin{equation}
\lambda= \begin{cases}1-\alpha \|(\sin \left(\frac{2 \pi t}{T}\right))\| \exp \left(-\frac{h-y}{d_0}\right), & 0 \leq \mathrm{t} \leq \mathrm{T} / 2, \\ 1-\alpha \|\sin \left(\frac{2 \pi t}{T}\right)\| \exp \left(-\frac{y}{d_0}\right), & \mathrm{T} / 2<\mathrm{t} \leq \mathrm{T},\end{cases}
\end{equation}
for 2D and 3D fish experiment. Here, $d_0$ controls the steepness of the decay, and we use $h/3$ in all experiments, $T$ denotes the period of contraction, $y$ denotes the material space coordinate in the principle axis of contraction, and $\alpha$ controls the strength of contraction. 

\begin{figure*}[t]
\centering
\begin{minipage}{.49\linewidth}
 \includegraphics[width=\linewidth]{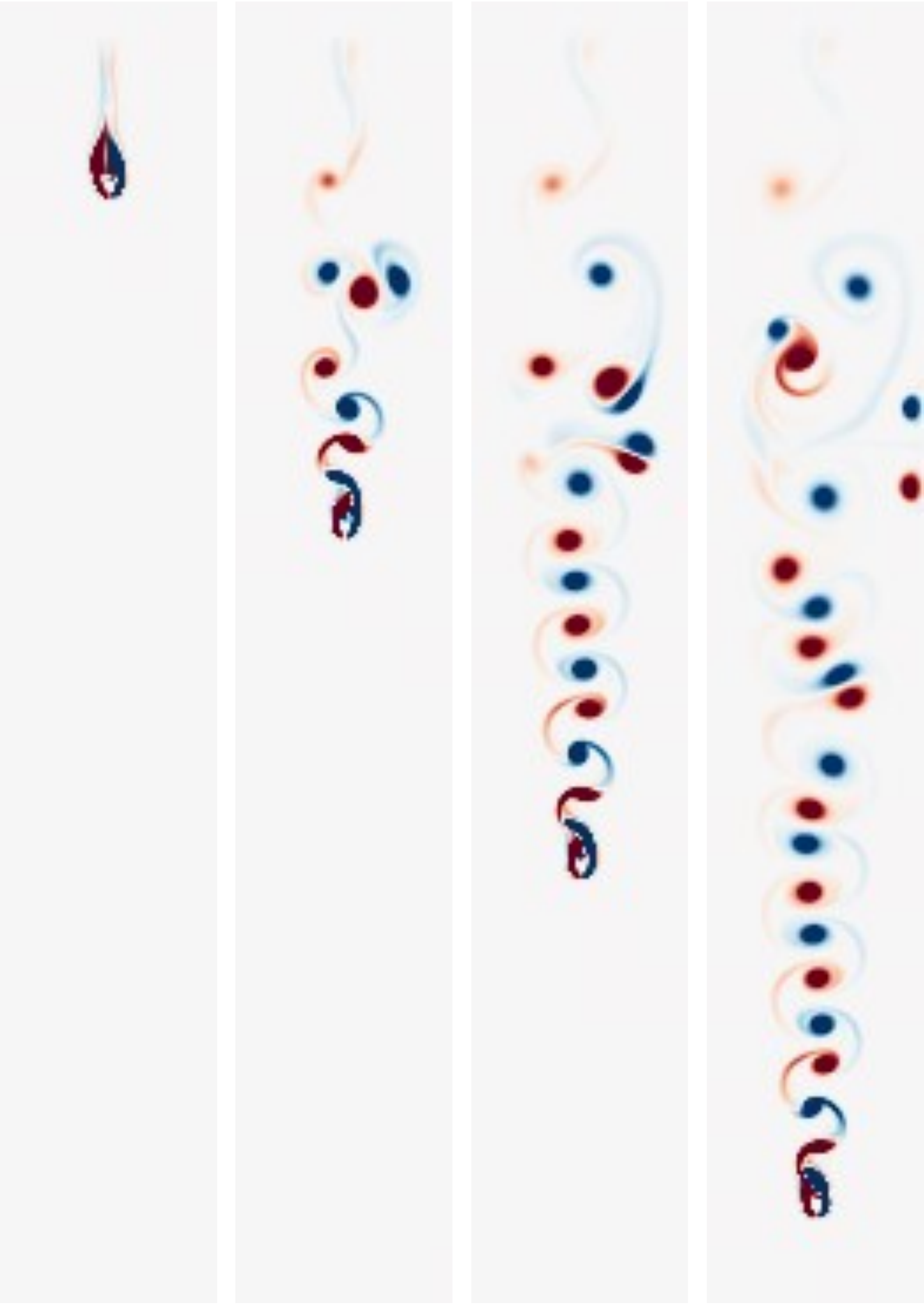}
 \caption{Vorticity plot of 2D cylinder sediment at time $t=1$, $3$, $5$, $7$}
 \label{fig:single_cyn}
\end{minipage}
\hspace{.01\linewidth}
\begin{minipage}{.49\linewidth}
  \includegraphics[width=\linewidth]{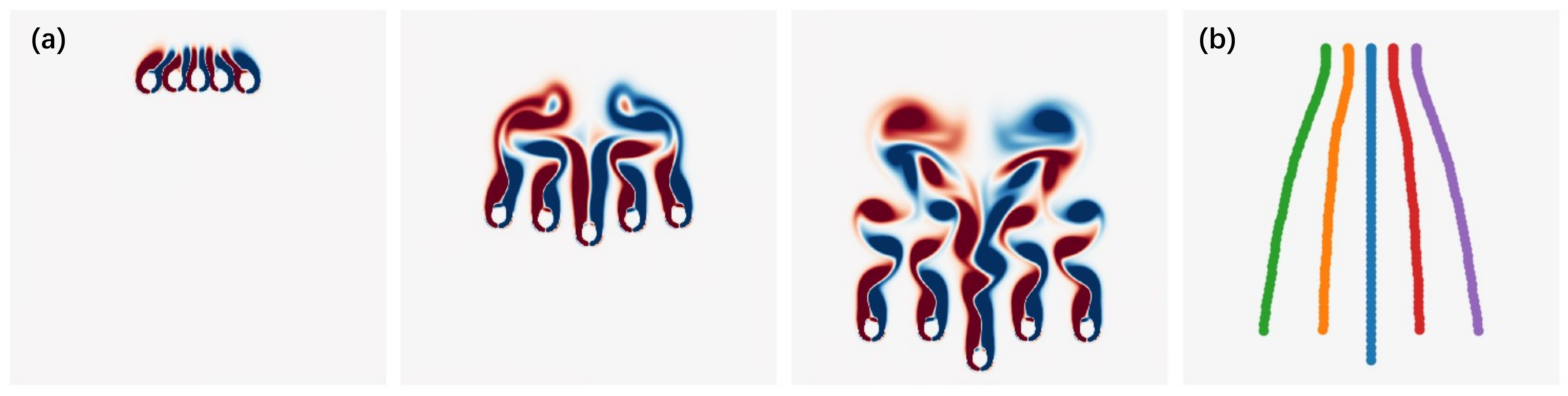}
 \caption{Figure (a) shows multi-cylinder sediment at time $t=0.15$, $0.45$, $0.65$. Figure (b) shows the trace for the center of mass of cylinders.}
 \label{fig:multi_cyn}
  \includegraphics[width=\linewidth]{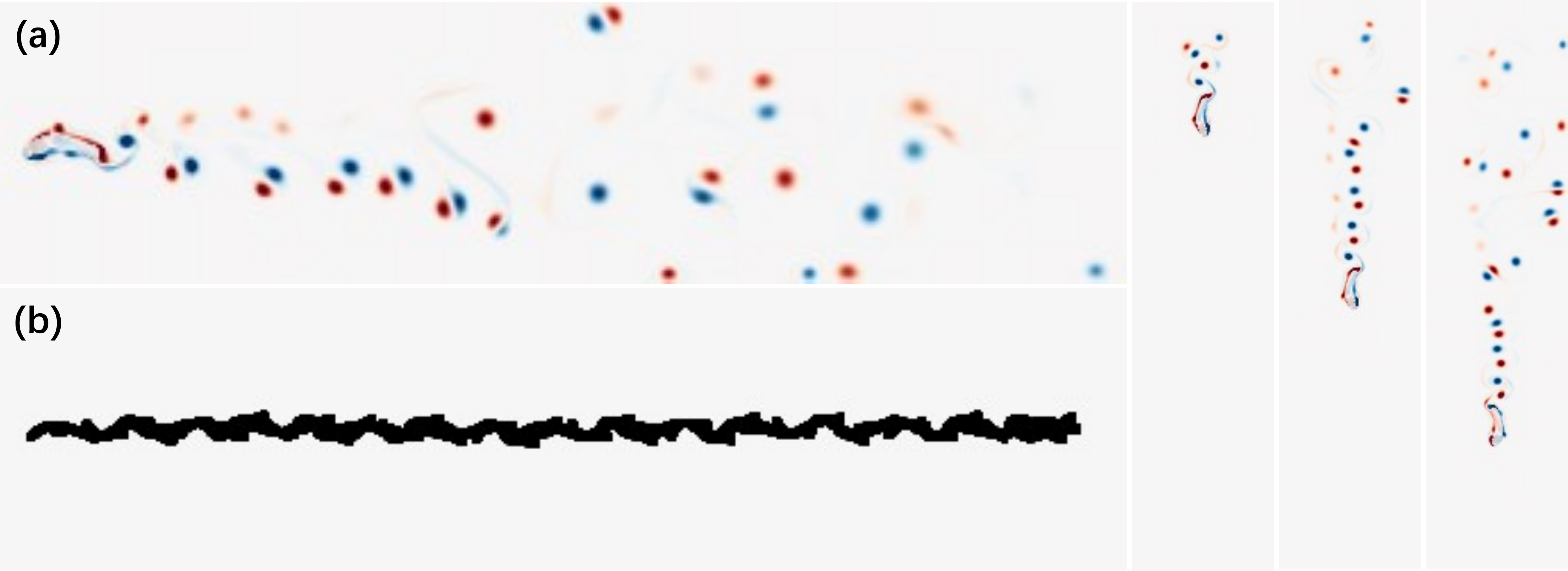}
 \caption{Figure (a) shows fish movement with vorticity at time $t=5$, $16$, $25$, $32$. Figure (b) shows the trace for fish movement.}
 \label{fig:2d_fish}
 \includegraphics[width=\linewidth]{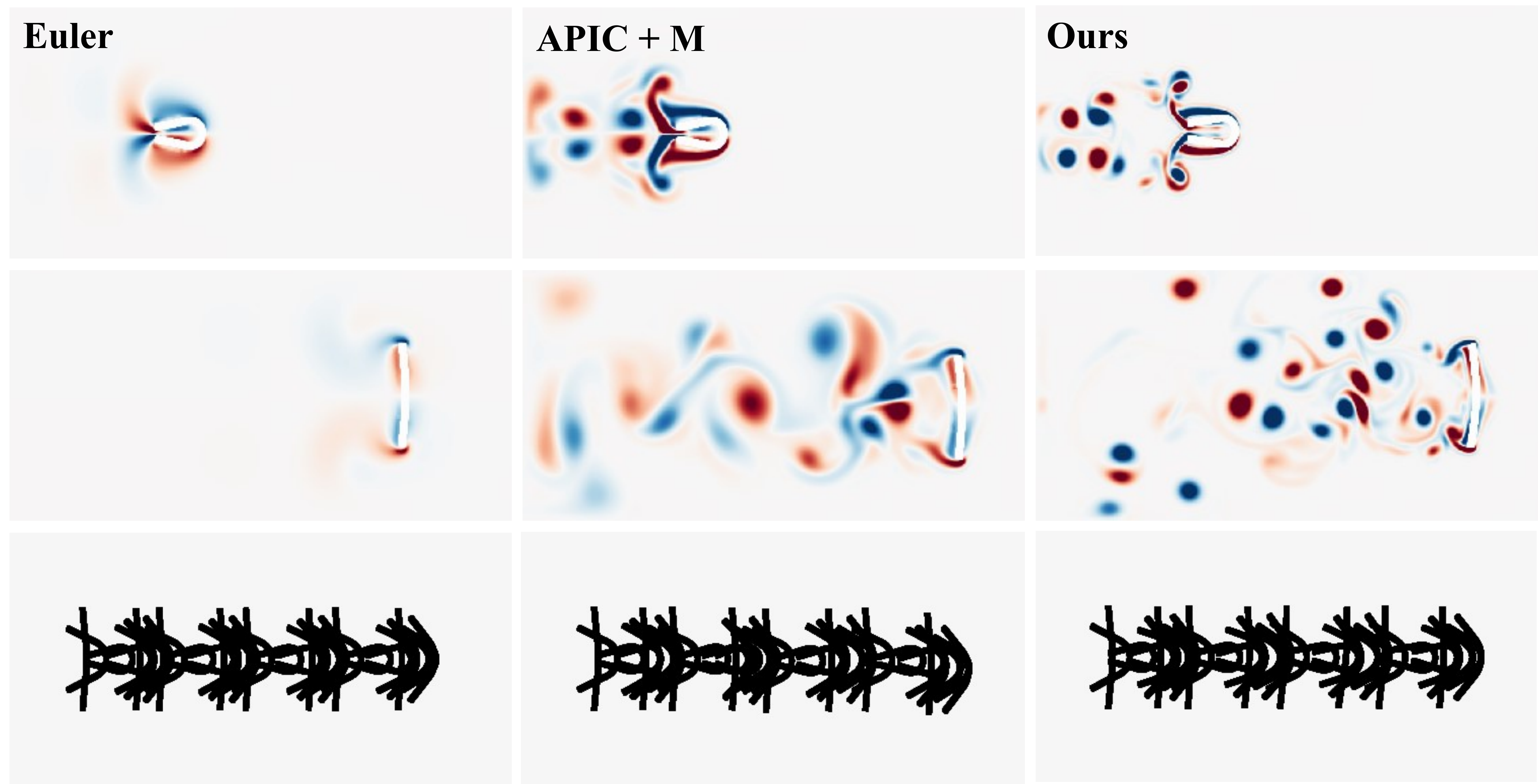}
 \caption{This experiment shows comparison of coupling between fluid and MPM using active strain method for self contractoin}
 \label{fig:2D_swimmer}
\end{minipage}
\end{figure*}

\paragraph{Comparison}
\label{sec:swim_result}
In this experiment, the active contraction area is set between 0.3 and 0.7 of the long axis of the solid. We use $\alpha=0.3$ and $T=2$, with a dynamic viscosity of $8 \times 10^{-6}$. Our method preserves the clearest vortical structures compared to the other two methods, as shown in Figure~\ref{fig:2D_swimmer}. We also show the swimming velocity of our swimer in Figure~\ref{fig:2D_fish_swimmer_vel}(a).


\subsection{2D / 3D Flag}
In 2D flag comparison, we compare our method with Euler and APIC+M by applying the IBM method combined with a mass-spring system for the flag. As shown in Figure~\ref{fig:2D_ibm_compare}, numerical viscosity dominates the fluid behavior in the Euler and APIC+M methods, resulting in a lack of interesting fluid vorticity. In contrast, our method encourages the formation of interesting vorticity structures. In the 3D flag experiment in Figure~\ref{fig:3D_cloth_compare}, no vortical structures are observed in the simulation produced by the Euler solver. Compared to APIC+M, our method displays clear and regular vorticity structures and coupled flapping motions.


\section{Examples}
In this section, we show various 2D/3D experiments to validate our coupling simulator. Table~\ref{tab:examples_table} shows the comprehensive list of examples. It is assumed that the shortest edge of the simulation domain is of unit length. All simulations were implemented with Taichi \cite{hu2019taichi} and performed on a laptop with an Intel i7-11800 H CPU and NVIDIA RTX 3080 GPU or a workstation equipped with AMD Ryzen Threadripper 5990X CPU and NVIDIA RTX 4090/A6000 GPUs. Depending on the GPU we used, our simulation takes an average of 1.7 seconds (E.g. 3D parachute or 3D Koinobori) to 5.2 seconds (E.g. 3D Fish or 3D Grass) for each step in 3D simulations.


\subsection{MPM Coupling}
The experiments below set the flow map reinitialization steps to 20 for 2D and 12 for 3D. The reinitialization step for narrowband fluid particles is set to 2 across all MPM coupling simulations.
\paragraph{2D cylinder sediment}
In the experiment shown in Figure~\ref{fig:single_cyn}, we show the result of a 2D cylinder falling from rest and converging to a constant speed due to the force from fluid acting on solid. The ratio between fluid and solid is set to 15:1, and gravity is set to 3. The radius of the cylinder is 0.03, and the center is initially placed at [0.16, 0.5]. Fluid viscosity is set to be $8\times10^{-5}$. We see Karman Vortex Street patterns created in this falling process as discussed in \cite{gazzola2011simulations}.

\paragraph{2D multi-cylinder}
As shown in Figure~\ref{fig:multi_cyn}, cylinders are placed at a distance of $3r$ between their centers, and their radius is $r=0.02$. The dynamic viscosity is set to be $6\times 10^{-4}$, the density ratio is 30:1, and gravity is 9.8. We observe symmetric patterns created for both vorticity and particles' paths.\cite{gazzola2011simulations}

\paragraph{2D Fish}
\label{sec:fish_result}
As discussed in \ref{sec:active_strain}, we use the active strain method to create self-contraction simulating muscle movement. In the experiment shown in Figur~\ref{fig:2d_fish}, we set the activation area between 0.6 and 0.9 in the x direction in the solid body's material space and principal contraction in the y direction. Dynamic viscosity is set to be $8\times10^{-6}$. The solid body has a dimension of $x=0.25$ and $y=0.01$. $T=2$ and $\alpha=0.25$ is used for contraction. The vorticity plot is shown in Figure~\ref{fig:2d_fish} and the swimming speed is shown in Figure~\ref{fig:2D_fish_swimmer_vel}(b).

\paragraph{3D Fish}
We use a setting similar to the 2D fish experiment. The solid has dimension $x=0.25$, $y=0.01$ and $z=0.1$. The principal contraction axis is the y direction and the activation area is between 0.6 and 0.9 in the x direction in the solid body's material space. $T=1.8$ and $\alpha=0.2$. The result is shown in Figure~\ref{fig:3d_fish}. We see a similar vortex structure created at the end of the tail compared to \cite{lin2019fluid}

\subsection{IBM Coupling}
In all the experiments below, we use reinitialization steps of 12 for fluid, and extended position-based dynamics (XPBD) is used to solve solid behaviors. In the 3D leaf and 3D parachute experiments, we adopt a suiting grid for tracking the solid center of mass to allow a longer falling path.

\paragraph{3D Grass}
We use multiple rectangular patches to form grass and fix their ending vertices at $y=0$. An incoming flow with variant velocity $v=0.2sin(2\pi t) + 0.1$ is used. Details of vortices are observed created from the surface of the grass during simulation. The results are shown in Figure~\ref{fig:3d_grass}.
\paragraph{3D Long Silk Flag}
An incoming flow of $v=0.1sin(2\pi t) + 0.2$ is applied with a random velocity between [0, 0.05]. Interestingly, tweaking and torsion of the flag are observed, and spiral patterns of the vortical structure are formed during simulation. See results in Figure~\ref{fig:3d_silk}.
\paragraph{3D Koinobori}
In Figure~\ref{fig:3d_cyn}, we create a cylinder mesh with a bigger open-up radius at the beginning compared to the end and connect it to a fixed point. We aim to use this shape in correspondence to Japanese Koinobori. A constant incoming flow is applied. 
\paragraph{3D Hair}
In this experiment, we fixed the hair ends and initialized it at a 45-degree angle. We applied the model in \cite{bender2014survey} for PBD hair simulation. We see hair flowing up by incoming flow, forming clear vortex structures as shown in Figure~\ref{fig:3d_hair}.

\revv{\paragraph{3D T-shirt}
Figure \ref{fig:3D_shirt} shows a shirt in a turbulent flow field to demonstrate the solid-fluid coupling with a complicated mesh topology. Surface wrinkles and turbulent wake vortices can be observed in the simulation.
}

\paragraph{3D Leaf \& 3D Parachute} In these two examples shown in Figure~\ref{fig:3d_leaf_para}, we use a body-suiting computation domain to allow a longer falling path to be tracked and simulated. We show the path of the falling object and the vorticity snapshots. For vorticity snapshots, we change the camera position by the center of mass of geometry for rendering. For more details, we refer readers to our supplemented video for world space rendering of the vorticity and falling process.

\paragraph{3D Combustion}
\revv{In Figure~\ref{fig:3d_fire} and \ref{fig:3d_grass_fire}}, we introduce combustion to the cylinder mesh depicted in Figure~\ref{fig:3d_cyn} and \revv{a wildfire scene on grass mesh in Figure~\ref{fig:3d_grass}}. The combustion source is initiated at a vertex of the mesh and spread across the entire mesh. We also impose a constant temperature and density field at the source, which is advected using the same fashion as PFM and visualized as fire and smoke, respectively. An incoming flow with a varying velocity is applied. Additionally, buoyancy is incorporated through $\beta(T - T_{\text{ambient}})$ where $\beta$, $T$, and $T_{\text{ambient}}$ represent the thermal expansion coefficient, temperature, and ambient temperature, respectively.

\section{Discussion}
\revv{We discuss the relation between our method and the previous impulse methods in the literature (e.g., \cite{nabizadeh2022covector, feng2022impulse, sancho2024impulse, zhou2024eulerianlagrangian, deng2023fluid}).}

\revv{
\paragraph{Impulse Stretching Stability} The current methods of simulating impulse fluid can be categorized into two categories: (1) evolving $\mathcal{F}$ and then solving $\mathcal{T}$ as its inverse \cite{feng2022impulse, sancho2024impulse}; (2) evolving $\mathcal{F}$ and $\mathcal{T}$ directly (without solving inverse) along a long-range flow map (e.g., on grid \cite{nabizadeh2022covector, deng2023fluid} or on particles \cite{zhou2024eulerianlagrangian, li2024lagrangian}). Impulse instability (i.e., a particle impulse magnifies itself due to a small numerical error, typically due to the stretching term, as reported in \cite{feng2022impulse, sancho2024impulse}) was observed when $\mathcal{T}$ was calculated as the $\mathcal{F}$ inverse. This instability was not observed when forward evolving $\mathcal{T}$ on a particle flow map (e.g. \cite{deng2023fluid, zhou2024eulerianlagrangian}). Our implementation follows the technical pathway of the second category by evolving $\mathcal{T}$ on particles. Therefore, we didn't employ extra treatment such as a Jacobian Limiter \cite{sancho2024impulse} to ensure the impulse numerical stability.
}

\revv{
\paragraph{Open Boundary Stability (CFL)}
We observed instability on the open boundary of the grid because $\mathcal{T}$ and $\mathcal{F}$ are undefined outside the domain. Unexpected fluctuating flow may enter the domain, limiting the CFL to a small number. This instability issue is closely related to the unsolved issue remaining in impulse simulations where the free surface / open boundary is present (e.g., \cite{li2024lagrangian,sancho2024impulse}) and remains an open problem to solve in future works. Such boundary instability does not occur for a wall boundary.
}

\revv{
\paragraph{Hessian Term in Flow Map}
As noted in recent literatures like \cite{sancho2024impulse}, the accurate advection of $\bm m$ involving particle-to-grid transfer requires the calculation of $\nabla \bm m$ whose transport equation depends on $\nabla \mathcal{T}$ and can be written as:
\begin{equation}
    \label{eq:evolve_grad_imp}
    \bm \nabla \bm m(\bm x,t) = \mathcal{T}^{T}_t\,\bm \nabla_{\psi} \bm{m}(\psi(\bm x),0)\,\mathcal{T}_t + \bm \nabla \mathcal{T}^{T}_t\,\bm{m}(\psi(\bm x),0).
\end{equation}
Such calculation (i.e., the calculation of $\nabla \mathcal{T}$) is challenging due to its high-order nature. However, we do not need to worry about this term because of the following reasons. Firstly and most importantly, in our method, by converting \(\bm{m}\) to \(\bm{u}\) first, the P2G transfer only requires \(\nabla \bm{u}\), eliminating the problematic calculation of \(\nabla \mathcal{T}\) from consideration. Secondly, in concurrent literature such as \cite{zhou2024eulerianlagrangian}, the authors propose to leave this term out of the calculation due to the observation that $\nabla \bm m$ mostly depends on the first term in Eq.~\ref{eq:evolve_grad_imp}.}

\begin{figure}
    \centering
    \includegraphics[width=0.99\columnwidth]{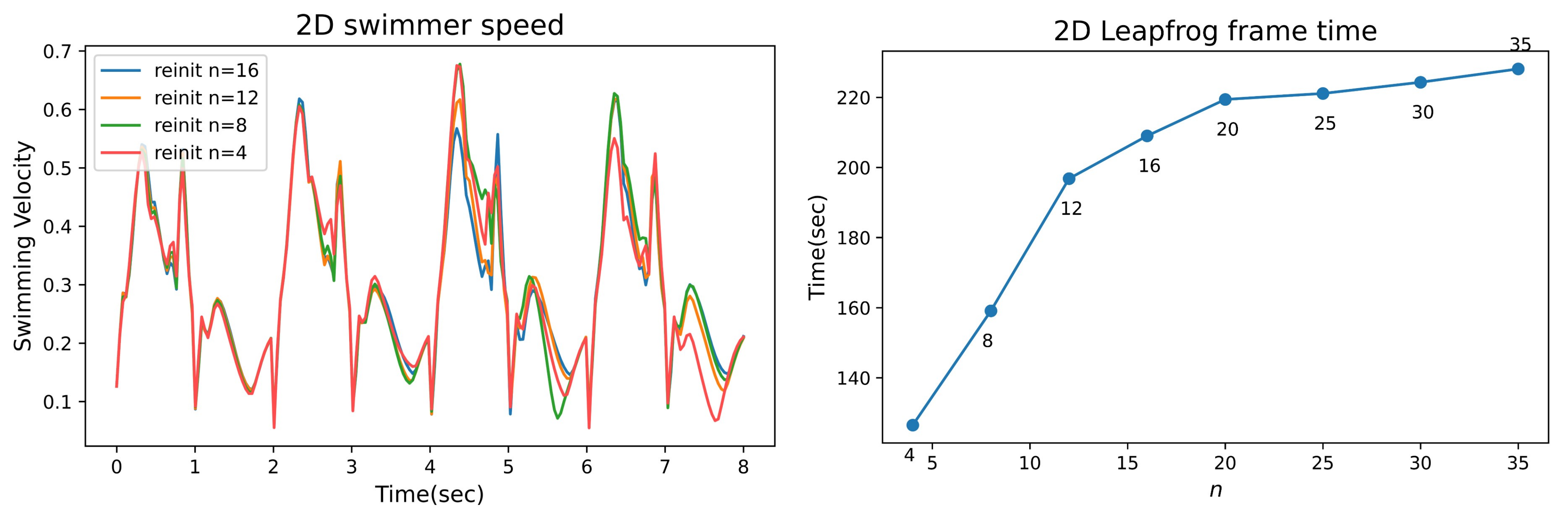}
    \caption{\revv{We study the influence of initialization frequency and the coupled solid swimming speed in the right picture and the influence of its vorticity preservation in a 2D leapfrog scenario. We $n$ to represent how many steps between reinitialization happens. We can see that solid speed is not largely influenced by this parameter. On the other hand, the vorticity preservation effect is not significantly improved after reinitialization frequency reaches 20.}}
    \label{fig:parameter}
\end{figure}

\revv{
\paragraph{Reinitialization Frequency} In Figure~\ref{fig:parameter}(a), we present the results of a parameter study investigating the influence of reinitialization frequency on solid velocity in solid-fluid coupling. We observe that the frequency of flow map reinitialization does not significantly impact the swimmer's speed. Nevertheless, reinitialization frequency is crucial for vorticity preservation, which is essential for smoke and soot visualization, as well as for managing multi-object and thin-shell object interactions (see Figure~\ref{fig:multi_cyn} and Figure~\ref{fig:3d_grass_fire}). Through experimenting with different reinitialization frequency parameters, we observe that vorticity preservation ability converges once a certain threshold is reached (See Figure~\ref{fig:parameter}(b)). As noted in \cite{nabizadeh2022covector}, excessively high reinitialization frequencies can lead to instability, particularly under conditions of substantial external force accumulation. Based on this observation, we have chosen to use a reinitialization frequency of \(n=20\) for 2D and \(n=12\) for 3D simulations instead of larger steps. For the number of particles per cell, we follow the standard practice in PFM, setting it to 16 for 2D and 8 for 3D, to obtain a balance between visual quality and GPU memory constraints.}

\section{Conclustion, Limitation, and Future Work}
In conclusion, we present a unified representation of solid and fluid dynamics using particle flow maps, where a single-step flow map is applied for solid simulation to integrate existing elastic body simulations and a longer flow map for fluids to preserve vortex structures. By coupling these two maps through an impulse-to-velocity transfer mechanism and managing force accumulation via particle path integrals, we have developed a robust coupling framework that adapts traditional solid-fluid coupling techniques to flow map models, as demonstrated in our MPM and IBM examples. 

The main limitation of our method is that the current framework relies on a weak coupling mechanism that depends on explicit force exchange between solid and fluid, rather than utilizing more sophisticated implicit momentum-conserving coupling schemes, such as the variational form \cite{batty2007fast} and monolithic projection \cite{robinson2008two}. These methods require a large system that implicitly includes the force exchange when formulating the system matrix. Integrating our method with these approaches would necessitate reformulating the two integrals in Eq.~\ref{eq:sfi_u_continuous} into an implicit system over the flow map process to accommodate monolithic coupling. This system reformulation presents an intriguing avenue for future work, particularly for enabling the coupling between complex vortical structures and stiff systems such as rigid bodies and articulated bodies within flow map methods. Specifically, we are motivated to further adapt our flow-map framework to projection-based immersed boundary methods (e.g., see \cite{guermond2006overview, taira2007immersed, wang2020monolithic, wang2015strongly}) to address the interactions between flow maps and solids with hard constraints.

Another limitation of our method is that our pipeline is confined to the framework of particle-based flow-map representation for the fluid component, which requires maintaining a large set of flow-map particles to buffer the coupling forces and transfer flow-map information between particles and the background grid for the Poisson solve. Reducing the cost of maintaining these particles could potentially decrease the computational costs for long-range flow maps. 

On another front, it would be interesting to explore purely Eulerian solid-fluid coupling methods based on flow maps (e.g., by extending the Eulerian solid-fluid interaction work \cite{teng2016eulerian} to a flow map framework), given the progress made with their Lagrangian counterparts. Additionally, our current system is focused on non-slip boundary conditions. It is unclear how to enforce other solid-fluid boundary conditions, such as non-penetration conditions, within a flow-map perspective. Motivated by progress in neighboring areas such as MPM (\cite{fang2020iq}), it would be valuable to explore different types of boundary conditions for solid-fluid coupling systems using flow maps.

Furthermore, our current approach does not support free-boundary flow. It would be intriguing to develop solid-fluid interaction frameworks that can simulate large-scale open-water phenomena (e.g., \cite{huang2021ships, Xiong2022Clebsch, wretborn2022guided}) with the flow-map-created vortical details around solid boundaries. The main challenge for the current approach is the difficulty of solving the impulse stretching terms for particles near the free surface (e.g., as pointed out by \cite{sancho2024impulse}). This issue remains a significant gap for flow-map methods due to the difficulty of handling impulse stretching near a free surface.

Overall, our future work aims to tackle these problems by facilitating an implicit formulation for two-way coupling between solids and free surfaces, and multi-phase flow map fluids with different types of boundary conditions.

\section*{Acknowledgements}
\revv{
We express our gratitude to the anonymous reviewers for their insightful feedback. We also thank Yuchen Sun, Taiyuan Zhang, and Shiying Xiong for their insightful discussion. Georgia Tech authors also acknowledge NSF IIS \#2433322, ECCS \#2318814, CAREER \#2433307, IIS \#2106733, OISE \#2433313, and CNS \#1919647 for funding support. Tao Du thanks Tsinghua University and Shanghai Qi Zhi Institute for their support. We credit the Houdini education license for video animations.
}

\begin{table*}
\centering\small
\begin{tabularx}{\textwidth}{Y | Y | Y | Y | Y | Y}
\hlineB{3}
Name & Figure & Resolution & CFL & Reinit. Steps of Flow Map & Particle Count Per Cell at Reinit. Step \\
\hlineB{2.5}
2D Cylinder Sediment & Figure~\ref{fig:single_cyn} & 764 $\times$ 128 & 0.5 & 20 & 16 \\
\hlineB{2}
2D Flag & Figure~\ref{fig:2D_ibm_compare} & 256 $\times$ 128 & 0.5 & 20 & 16 \\
\hlineB{2}
2D Multi-Cylinder Sediment & Figure~\ref{fig:multi_cyn} & 256 $\times$ 256 & 0.5 & 20 & 16 \\
\hlineB{2}
2D Fish & Figure~\ref{fig:2d_fish} & 512 $\times$ 128 & 0.5 & \revv{20} & 16 \\
\hlineB{2}
2D Flapping Swimmer & Figure~\ref{fig:2D_swimmer} & 256 $\times$ 128 & 0.5 & \revv{20} & 16 \\
\hlineB{2}
\revv{3D Smoke Plume} & \revv{Figure~\ref{fig:3d_bouyancy}} & \revv{256 $\times$ 128 $\times$ 128} & \revv{0.5} & \revv{12} & \revv{8} \\
\hlineB{2}
3D Fish & Figure~\ref{fig:3d_fish} & 256 $\times$ 128 $\times$ 128 & 0.5 & 12 & 8 \\
\hlineB{2}
3D Flag & Figure~\ref{fig:3D_cloth_compare} & 256 $\times$ 128 $\times$ 128 & 0.25 & 12 & 8 \\
\hlineB{2}
3D Parachute & Figure~\ref{fig:3d_leaf_para} & 256 $\times$ 128 $\times$ 128 & 0.25 & 12 & 8 \\
\hlineB{2}
3D Leaf & Figure~\ref{fig:3d_leaf_para} & 256 $\times$ 128 $\times$ 128 & 0.25 & 12 & 8 \\
\hlineB{2}
3D Koinobori & Figure~\ref{fig:3d_cyn} & 256 $\times$ 128 $\times$ 128 & 0.25 & 12 & 8 \\
\hlineB{2}
3D Grass & Figure~\ref{fig:3d_grass} & 256 $\times$ 128 $\times$ 128 & 0.25 & 12 & 8 \\
\hlineB{2}
3D Hair & Figure~\ref{fig:3d_hair} & 256 $\times$ 128 $\times$ 128 & 0.25 & 12 & 8 \\
\hlineB{2}
3D Long Silk Flag & Figure~\ref{fig:3d_silk} & 256 $\times$ 128 $\times$ 128 & 0.25 & 12 & 8 \\
\hlineB{2}
3D Koinobori Combustion & Figure~\ref{fig:3d_fire}& 256 $\times$ 128 $\times$ 128 & 0.25 & 12 & 8 \\
\hlineB{2}
\revv{3D Grass Combustion} & \revv{Figure~\ref{fig:3d_grass_fire}} & \revv{256 $\times$ 128 $\times$ 128} & \revv{0.25} & \revv{12} & \revv{8} \\
\hlineB{2}
\revv{3D T-shirt} & \revv{Figure~\ref{fig:3D_shirt}} & \revv{256 $\times$ 128 $\times$ 128} & \revv{0.25} & \revv{12} & \revv{8} \\
\hlineB{3}
\end{tabularx}
\vspace{5pt}
\caption{The catalog of all our 2D and 3D simulation examples.}
\label{tab:examples_table}
\end{table*}


{
\bibliography{ref}
\bibliographystyle{ACM-Reference-Format}
}

\appendix
\section{Implementation Details}
\subsection{Implementation detail for MPM}
We explain our implementation for MPM solid substeps and removing fluid particles in solid regions. Solid substeps need to be coupled with fluid, and fluid particles in solid regions must be removed during reinitialization for flow maps and fluid particles. Therefore, we choose to use narrowband particles resampled at a higher frequency than the fluid particles in distant regions to facilitate coupling at solid substeps with fluid and to remove fluid particles in solid regions.
\label{app:mpm_sup}
\subsubsection{Narrowband fluid particles \& Resampling}
\label{sec:resample}
\begin{figure}
    \centering
    \includegraphics[width=0.99\columnwidth]{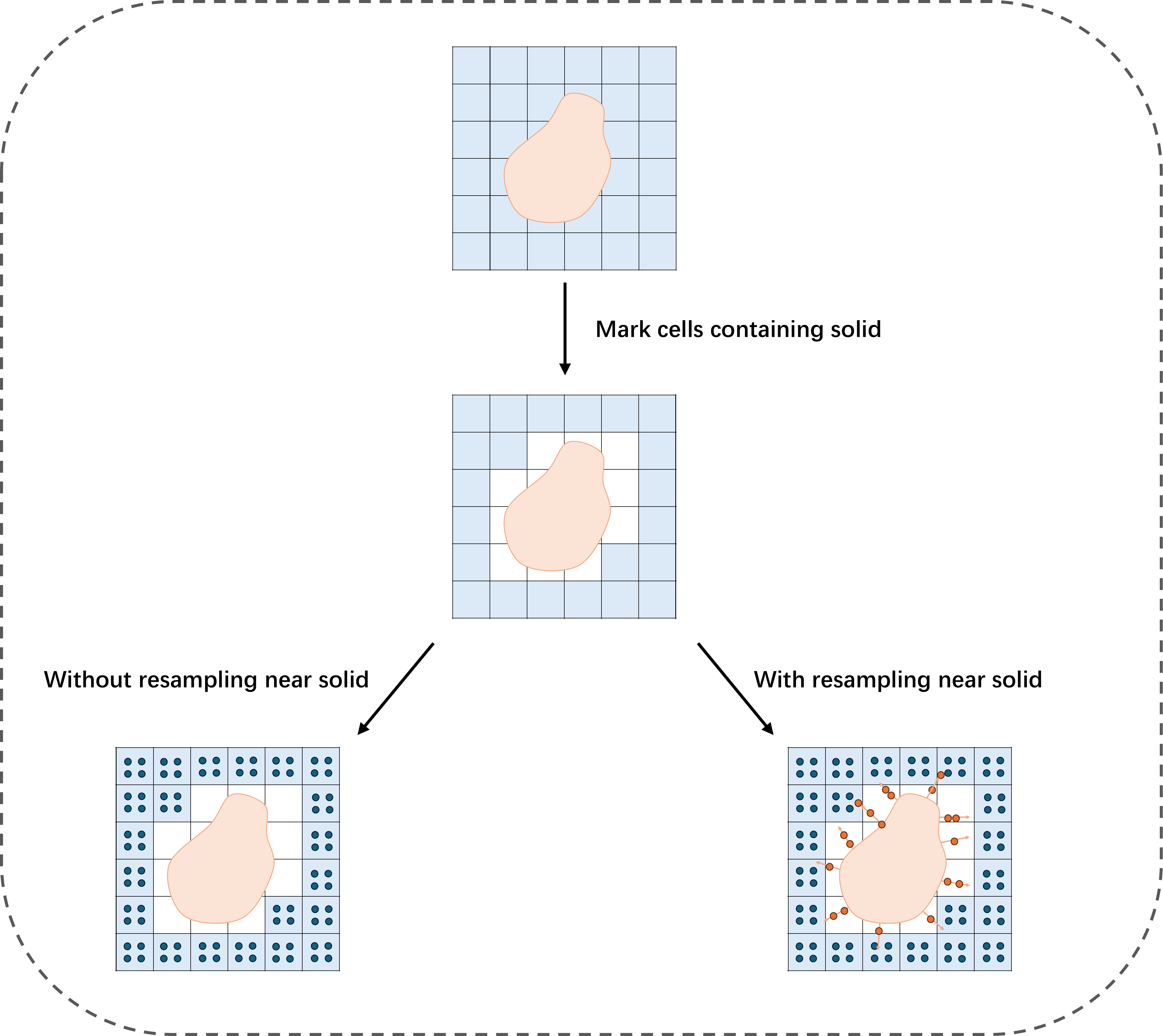}
    \caption{We show that if we do not perform any resampling in the cells where fluid particles are removed, an empty band that contains no fluid particles will be created. And by using our resampling method, we can fill in this band easily.}
    \label{fig:resample}
\end{figure}
In order to quickly remove particles in the solid region without the need to reconstruct the level set from particles at each step \cite{boyd2012multiflip}, we utilize the deformation gradient carried by that are particles that are sampled exactly on a solid surface. Directly removing particles in every cell containing solid particles would create a gap between solid and fluid particles, as illustrated in Figure~\ref{fig:resample}. Therefore, we need to find a method to resample fluid particles near the solid surface to fill in this gap.

Noticing that normal vectors can also be advected through the flow map according to the following relationship:
\begin{equation}
    \bm n_c = \mathcal{T}_{[0, c]}^{T}\bm{n}_0
\end{equation}
we sample on the solid surface before the simulation begins and document the surface points as $\bm x^{\text{surface}}$ and their initial surface normals as $\bm n^{\text{surface}}_0$. A random distance $D \in (0, 1.5\Delta x]$ is recorded for each $\bm x^{\text{surface}}$ before the simulation begins, and the resampling during each reinitialization step of narrowband fluid particles $\bm{x}^n$ naturally becomes:
\begin{equation}
    \bm{x}^n = \bm x^{\text{surface}} + (\mathcal{T}^{\text{surface}}_{[0, c]})^{T}\bm{n}^{\text{surface}}_0 D
\end{equation}
We outline the algorithm for resampling in Alg.~\ref{alg:sampling}.

\begin{algorithm}[H]
\caption{Sampling Method}
\label{alg:sampling}
\begin{algorithmic}[1]
\State Random sample points on surface and document as $\bm x^{\text{surface}}$ and their normals as $\bm n^{\text{\revv{surface}}}_0$;
\State Perturb particle on surface along their normal $\bm n^{\text{surface}}_0$ with distance $D$;
\State Update $\bm x^{\text{surface}}$ and $\mathcal{T}^{\text{surface}}$ in simulation;
\State At $t$ timestep, $\bm{x}^n = \bm x^{\text{surface}} + (\mathcal{T}^{\text{surface}}_{[0, c]})^{T}\bm{n}^{\text{surface}}_0 D$;
\end{algorithmic}
\end{algorithm}

\begin{algorithm}[t]
\caption{Reinit for coupling MPM}
\label{alg:reinit}
\begin{algorithmic}[1]

\State $j \gets k \Mod {n^L}$;
\State $h \gets k \Mod {n^S}$;
\If{$j$ = 0}
\State Uniformly distribute particles;
\State Mask out particles in solid
\State Reinitialize $\bm m^f$ for all fluid particles $\bm x^f$; 
\State Reinitialize $\mathcal{T}^f$, $\mathcal{F}^f$ to identity\\; 
\State Empty stored pressure correction buffer $\bm{\Lambda}^f$
\State Empty stored external force buffer $\bm{\Upsilon}^f$
\EndIf

\If{$h$ = 0}
\State Resample near solid particles using solid normal
\State Reinitialize $\bm m^n$ for all near solid fluid particles $\bm x^n$; 
\State Reinitialize $\mathcal{T}^n$, $\mathcal{F}^n$ to identity\\; 
\State Empty stored pressure correction buffer $\bm{\Lambda}^n$
\State Empty stored external force buffer $\bm{\Upsilon}^n$

\EndIf

\end{algorithmic}
\end{algorithm}

Subsequently, the resampled narrowband fluid particles will always remain outside the solid, and we can safely remove fluid particles from every cell that contains solid particles at the reinitialization step.

\subsubsection{Narrowband fluid particles in solid substeps}

Based on the assumption that minimal movement will occur within solid substeps for fluid, we only utilize narrowband fluid particles for P2G and G2P during these substeps. Additionally, we use the $\mathcal{F}^n$ and $\mathcal{T}^n$ advected through solid substeps to update the impulse $\bm m^n$ and calculate $\bm \Lambda^n$ and $\bm \Upsilon^n$ when synchronizing solid and fluid states for coupling. Incompressibility is enforced by solving the Poisson equation during the synchronization step, rather than using the artificial bulk modulus cited in \cite{hu2018moving}.

\subsubsection{Pseudo Code for Coupling with MPM}
\label{sec:mpm_app}
Below we show the full pseudo-code for coupling impulse fluid with MPM under the time integration scheme we proposed in Algorithm~\ref{alg:general_pipeline}. Limited modifications are required to adapt MPM into the pipeline as shown below.

\begin{algorithm}[H]
\caption{Coupling with MPM}
\label{alg:mpm_coupling}
\begin{algorithmic}[1]
\For{$k$ in total steps}
\State Do reinit if needed based on Alg.~\ref{alg:reinit}

\State Compute $\Delta t$ with $\bm{u}_i$ and the CFL number;
\State Compute $\Delta t^s$ with solid parameter and the sound CFL;
\State Decide the number of substeps for solid

\State Estimate $\bm u_{\text{mid}}$ with Alg.~\ref{alg:midpoint_mpm}

\State March $\bm x^f_c$, $\mathcal{T}^f_{[a, c]}$ $\mathcal{F}^f_{[c, a]}$ with $\bm u_{\text{mid}}$ and $\Delta t$; 
\State March the other half MPM substep with Alg.~\ref{alg:mpm_substep} 

\State G2P $\bm u_{sub}$ to get $\bm u^s_c$

\State Get $\bm m^f_c$ and $\bm m^n_c$ using Eq.~\ref{eq:update_equation}
\State  Compute $\bm u_c^{f*}$ using $\bm{\Lambda}^f_b$, $\bm{\Upsilon}^f_b$, $\bm{u}_{\text{mid}}$, $\bm m^f_c$ with Eq.~\ref{eq:update_equation}.

\State  Compute $\bm u_c^{n*}$ using $\bm{\Lambda}^n_b$, $\bm{\Upsilon}^n_b$, $\bm{u}_{\text{mid}}$, $\bm m^n_c$ with Eq.~\ref{eq:update_equation}.

\State Compute $\nabla \bm u^f_c$ $\nabla \bm u^n_c$ using $\bm u_{\text{mid}}$

\State Compute $\bm u_i$ by P2G using $\bm x^f_c$, $\bm x^n_c$, $\bm x^s_c$, $\bm u^f_c$, $\bm u^n_c$, $\bm u^s_c$, $\nabla \bm u^f_c$, $\nabla \bm u^n_c$, $\mathcal{F}^s$
\State Add gravity on $\bm u_i$ if needed
\State Add viscosity on $\bm u_i$ if needed

\State P2G density carried by fluid and solid particles
\State Solve Poisson

\State Update $\bm \Upsilon^f_c$, $\bm \Upsilon^n_c$ by adding external force to buffer following Eq.~\ref{eq:update_equation}. 
\State Update $\bm \Lambda^f_c$, $\bm \Lambda^n_c$ by adding pressure correction to buffer following Eq.~\ref{eq:update_equation}. 

\EndFor{}
\end{algorithmic}
\end{algorithm}

\begin{algorithm}
\caption{Midpoint MPM}
\label{alg:midpoint_mpm}

\begin{algorithmic}[1]
\State March half MPM substep with Alg.~\ref{alg:mpm_substep}
\State March $\bm x^f$ with $\bm{u}_i$ and $0.5\Delta t$ and get $\bm u^f$, $\nabla \bm u^f$
\State G2P $\bm u_{sub}$ to $\bm u^n$ and $\bm u^s$ and compute $\nabla \bm u^n$
\State P2G using $\bm x^f$, $\bm x^n$, $\bm x^s$, $\bm u^f$, $\bm u^n$, $\bm u^s$, $\nabla \bm u^f$, $\nabla \bm u^n$, $\mathcal{F}^s$
\State P2G density carried by fluid and solid particles
\State Solve Poisson and get $\bm u_{\text{mid}}$

\end{algorithmic}
\end{algorithm}

\begin{algorithm}
\caption{MPM substep}
\label{alg:mpm_substep}
\begin{algorithmic}[1]

\State $\bm u_{sub} \gets \bm u$
\For{substeps}
\State march $\bm x^n$, $\bm x^s$, $\mathcal{T}^n$, $\mathcal{F}^n$, $\mathcal{T}^s$, $\mathcal{F}^s$ with $\bm u_{sub}$ 
\State G2P $u_{sub}$ to $\bm u^n$ and $\bm u^s$ and compute $\nabla \bm u^n$
\State P2G using $\bm x^n$, $\bm x^s$, $\bm u^n$, $\bm u^s$, $\nabla \bm u^n$, $\mathcal{F}^s$
\EndFor{}

\end{algorithmic}
\end{algorithm}

\subsection{Implementation detail for IBM}
\label{app:ibm_sup}
\subsubsection{Marching solid by XPBD} 
Except for the 2D validation tests, where we used a mass-spring system for our solid model, we employ XPBD to simulate solids for better visual effects. We applied the edge length and bending constraints as outlined in \cite{bender2014survey}. We used an iteration count of 50 for solving XPBD constraints in all our simulations, with solid $\Delta t_{\text{solid}}$ set to 0.0005 for stability.

One caveat to note is that we use substeps for solid calculations. Therefore, the force that is spread to the grid velocity is calculated in an explicit fashion, as follows:
\begin{equation}
    \bm f = \rho\frac{\bm u_{c} - \bm u_{b}}{\Delta t}
\end{equation}

\begin{algorithm}[t]
\caption{Reinit for coupling MPM}
\label{alg:reinit}
\begin{algorithmic}[1]

\State $j \gets k \Mod {n^L}$;
\State $h \gets k \Mod {n^S}$;
\If{$j$ = 0}
\State Uniformly distribute particles;
\State Mask out particles in solid
\State Reinitialize $\bm m^f$ for all fluid particles $\bm x^f$; 
\State Reinitialize $\mathcal{T}^f$, $\mathcal{F}^f$ to identity\\; 
\State Empty stored pressure correction buffer $\bm{\Lambda}^f$
\State Empty stored external force buffer $\bm{\Upsilon}^f$
\EndIf

\If{$h$ = 0}
\State Resample near solid particles using solid normal
\State Reinitialize $\bm m^n$ for all near solid fluid particles $\bm x^n$; 
\State Reinitialize $\mathcal{T}^n$, $\mathcal{F}^n$ to identity\\; 
\State Empty stored pressure correction buffer $\bm{\Lambda}^n$
\State Empty stored external force buffer $\bm{\Upsilon}^n$

\EndIf

\end{algorithmic}
\end{algorithm}

\begin{algorithm}[H]
\caption{Reinit for coupling IBM}
\label{alg:reinit_ibm}
\begin{algorithmic}[1]

\State $j \gets k \Mod {n^L}$;
\If{$j$ = 0}
\State Uniformly distribute particles;
\State Reinitialize $\bm m_s$ for all fluid particles $\bm x^f$; 
\State Reinitialize $\mathcal{T}^f$, $\mathcal{F}^f$ to identity\\; 
\State Empty stored pressure correction buffer $\bm{\Lambda}^f$
\State Empty stored external force buffer $\bm{\Upsilon}^f$
\EndIf
\end{algorithmic}
\end{algorithm}

\subsubsection{Pseudo Code for Coupling with IBM}
\label{sec:ibm_app}
Below we show the full pseudo-code for coupling covector/impulse fluid with IBM under the general pipeline we proposed in Algorithm~\ref{alg:general_pipeline}. The only modification required is to use XPBD to solve for solid behaviors.

\begin{algorithm}[H]
\caption{Coupling with IBM}
\label{alg:ibm_coupling}
\begin{algorithmic}[1]
\For{$k$ in total steps}
\State Do reinitialization if needed based on Alg.~\ref{alg:reinit_ibm}

\State Compute $\Delta t$ with $\bm{u}_i$ and the CFL number;
\State Determine $\Delta t^s$;
\State Decide the number of substeps for solid

\State Estimate $\bm u_{\text{mid}}$ with Alg.~\ref{alg:midpoint_ibm}

\State March $\bm x^f$, $\mathcal{T}^f_{[a, c]}$ $\mathcal{F}^f_{[c, a]}$ with $\bm u_{\text{mid}}$ and $\Delta t$; 
\State March the other half IBM substep with Alg.~\ref{alg:ibm_substep} 

\State Get $\bm m_c$ using Eq.~\ref{eq:update_equation}
\State  Compute $\bm u_c^{f*}$ using $\bm{\Lambda}_b$, $\bm{\Upsilon}_b$, $\bm{u}_{\text{mid}}$, $\bm m_c$ with Eq.~\ref{eq:update_equation}.

\State Compute $\nabla \bm u^f_c$ using $\bm u_{\text{mid}}$

\State Compute $\bm u_i$ by P2G using $\bm x^f_c$, $\bm u^f_c$, $\nabla \bm u^f_c$
\State Add gravity on $\bm u_i$ if needed
\State Add viscosity on $\bm u_i$ if needed

\State Compute force $\bm f^s$ with Update $\rho\frac{(\bm u^s_{c} - \bm u^{s}_b)}{\Delta t}$

\State Spread $\bm f^s$ with IBM kernel to $\bm u_i$
\State Solve Poisson

\State Get $\bm f^f$ on fluid particles using G2P

\State Update $\bm \Upsilon_c$ by adding external force to buffer as Eq.~\ref{eq:update_equation}. 
\State Update $\bm \Lambda_c$ by adding pressure correction to buffer as Eq.~\ref{eq:update_equation}. 

\EndFor{}
\end{algorithmic}
\end{algorithm}

\begin{algorithm}
\caption{Midpoint IBM}
\label{alg:midpoint_ibm}

\begin{algorithmic}[1]
\State Get solid velocity using IBM kernel and $\bm{u}_i$
\State March $\bm x^s$ with Alg.~\ref{alg:ibm_substep}

\State Compute force $\bm f^s$ with Update $\rho\frac{(\bm u_{b+0.5\Delta t} - \bm u_{b})}{\Delta t}$

\State Get $\bm u^*_{\text{mid}}$ with RK4 semi-Lagragian update.
\State Spread force with IBM kernel to $\bm u^*_{\text{mid}}$
\State Solve Poisson to get $\bm u_{\text{mid}}$
\end{algorithmic}
\end{algorithm}

\begin{algorithm}[H]
\caption{IBM Solid substep}
\label{alg:ibm_substep}
\begin{algorithmic}[1]

\For{substeps}
\If {With XPBD}
\State Update $\alpha$ based on $\Delta t$ for each $\mathbb{C}$
\State Predict solid location $\bm x^s$ with forward Euler
\State Solve with XPBD iterations and get updated $\bm x^s$
\State Update $\bm u^s$
\Else
\State Explicit Euler solving spring-mass system to update $\bm x^s$
\EndIf 
\EndFor{}

\end{algorithmic}
\end{algorithm}
\end{document}